\documentclass[]{aa}  

\usepackage{graphicx}
\usepackage{txfonts}
\usepackage[backref, pagebackref=false, hyperindex=true, breaklinks=true, colorlinks=true, urlcolor=magenta, linkcolor=magenta, citecolor=NavyBlue, citebordercolor={0 1 0}, pagecolor=red, bookmarks=true, filecolor=blue,bookmarksopen=true, plainpages=false, pdfpagemode=UseThumbs]{hyperref}
\usepackage[dvipsnames]{xcolor}
\usepackage{ulem}
\usepackage{siunitx}
\DeclareSIUnit\msun{M_\odot}
\DeclareSIUnit\year{yr}
\DeclareSIUnit\au{AU}
\DeclareSIUnit\eV{eV}
\DeclareSIUnit\erg{erg}

\defcitealias{Weder2023}{WME23}
\def\WME23{\citetalias{Weder2023}}

\usepackage[textsize=tiny]{todonotes}
\usepackage{xcolor}
\usepackage{multicol}
\usepackage{subcaption}

\begin{document}

\title{Inferring the physics of protoplanetary disc evolution from the irradiated Cygnus OB2 region}
\subtitle{A comparison of viscous and MHD wind-driven scenarios }

\author{Jesse Weder \inst{\ref{unibe}}
        \and
        Andrew J. Winter \inst{\ref{andrew_affiliation_1}, \ref{andrew_affiliation_2}, \ref{andrew_affiliation_3}}
        \and
        Christoph Mordasini \inst{\ref{unibe},\ref{CSH}}
        }

\institute{
        Division of Space Research and Planetary Sciences, Physics Institute, University of Bern, Gesellschaftsstrasse 6, 3012 Bern, Switzerland\\ \email{jesse.weder@unibe.ch} \label{unibe}
        \and
        Center for Space and Habitability, University of Bern, Gesellschaftsstrasse 6, 3012 Bern, Switzerland\label{CSH}
        \and
        Max-Planck Institute for Astronomy (MPIA), Königstuhl 17, 69117 Heidelberg, Germany \label{andrew_affiliation_1} 
        \and
        Université Côte d'Azur, Observatoire de la Côte d'Azur, CNRS, Laboratoire Lagrange, 06300 Nice, France  \label{andrew_affiliation_2}
        \and    
        Astronomy Unit, School of Physics and Astronomy, Queen Mary University of London, London E1 4NS, UK
        \label{andrew_affiliation_3}
        }

\date{Received 9 July 2025; accepted 29 October 2025}

% \abstract{}{}{}{}{}
% 5 {} token are mandatory
\abstract
{The physical processes driving protoplanetary disc evolution is of paramount importance for understanding planet formation. Our current understanding has crystallized around two possible evolution scenarios (turbulent viscosity and magnetohydrodynamic (MHD) wind-driven). Which of these processes dominates remains unclear.}
{Our aims are twofold: Firstly, we investigate whether a single set of model parameters can reproduce the observational constraints of non-irradiated and irradiated discs. Secondly, we propose a novel approach to break degeneracies between these two scenarios by studying the relation of stellar accretion rate and externally driven wind mass-loss rates, which evolve differently depending on the mechanism of angular momentum transport in the outer disc and test this approach using our models.}
{We simulate the evolution of synthetic populations of protoplanetary discs using 1D vertically integrated models for both viscous and MHD wind-driven disc evolution including both internal X-ray and external far ultraviolet (FUV) photoevaporation for both evolution scenarios. We investigate both weak and strong FUV field environments, where the strong FUV field is calculated based on an environment similar to the Cygnus OB2 association. We study the time evolution of the disc fraction, disc mass - stellar accretion rate relation, the spatial variation of the disc fraction in a highly irradiated cluster, the evolution of disc radii, and the evolution of accretion rates versus wind-mass-loss rates.}
{While both evolution scenarios are capable of reproducing observational constraints, our simulations suggest that different parameters are needed for the angular momentum transport to explain disc lifetimes and disc mass - stellar accretion rate relation in  weakly and strongly irradiated regions. We find that the predicted median disc radii are much larger in low FUV environments compared to Cygnus OB2, but also decreasing with time. In the viscous scenario, the median disc radius in a low FUV field environment is $\sim100\,\mathrm{au}$ larger than for the MHD wind-driven scenario. We further demonstrate that studying stellar accretion rates and externally driven wind mass-loss rates (provided that they can be isolated from internally driven winds; i.e. MHD wind) is indeed a promising way of disentangling the two evolution scenarios.}
{The fact that not a single set of parameters for angular momentum transport is able to reproduce disc lifetimes in both low and highly irradiated regions at the same time indicates a fundamental difference in these two regions.}


\maketitle
\nolinenumbers

%===============
\section{Introduction}
%===============
Circumstellar discs are a direct consequence of star formation and the understanding of the evolution and dispersal is fundamental for understanding planet formation. A critical aspect of global disc evolution is how angular momentum is redistributed through the disc to sustain observed accretion rates \citep[e.g.][]{Manara2022}. One scenario involves a substantial turbulent velocity field, driven by some hydrodynamic instability, resulting in an effective turbulent viscosity \citep{Lust1952,Shakura1973,LyndenBell1974}. As observational constraints do not support high levels of turbulence \cite[see review by][]{pinte_kinematic_2023}, the magnetohydrodynamic (MHD) wind-driven scenario has gained popularity recently. In this scenario, the angular momentum is removed by gas flowing along large scale magnetic field lines that are attached to the disc \citep{Blandford1982,Konigl2010}.

Numerous efforts have been made to find the dominant process driving disc evolution \cite[e.g.][]{Lodato2017,Trapman2020,trapman_effect_2022,Manara2022,Tabone2022b,alexander_distribution_2023,somigliana_time_2023}. Most recently first results from the ALMA Survey of Gas Evolution of PROtoplanetary Disks (AGE-PRO) have been released \cite{zhang_alma_2025}. The aim of this survey was to systematically characterise discs in nearby star forming regions of different age to nail down disc evolution processes. \cite{tabone_alma_2025} used a population syntheses approach to interpret the results of AGE-PRO with both viscous and MHD wind-driven disc evolution models, favouring MHD wind-driven models. However, broadly speaking, these studies have not been conclusive, owing both to parameter degeneracies and to complications in mapping the physical disc properties to observations \citep[e.g.][]{miotello_setting_2023}.

Further complicating matters, protoplanetary discs are not necessarily evolving driven by a single process in isolation but by a suite of processes occurring at different times and locations. These processes may be internal in nature (involving only the closed star-disc system), however many of the processes that are increasingly understood to have a substantial impact on disc evolution are external. As stars do not form in isolation, but in clusters of stars, external photoevaporation driven by far ultraviolet (FUV) radiation \citep{Winter_Haworth_2022},  dynamic encounters between stars \citep{Cuello2023}, and infall from the interstellar medium \citep{kuffmeier_episodic_2018} may be important for global disc evolution. These processes can be complicating factors, but taking them into account is not only necessary, but may also help to break degeneracies. 

\begin{figure*}
    \centering
    \includegraphics[width=15cm]{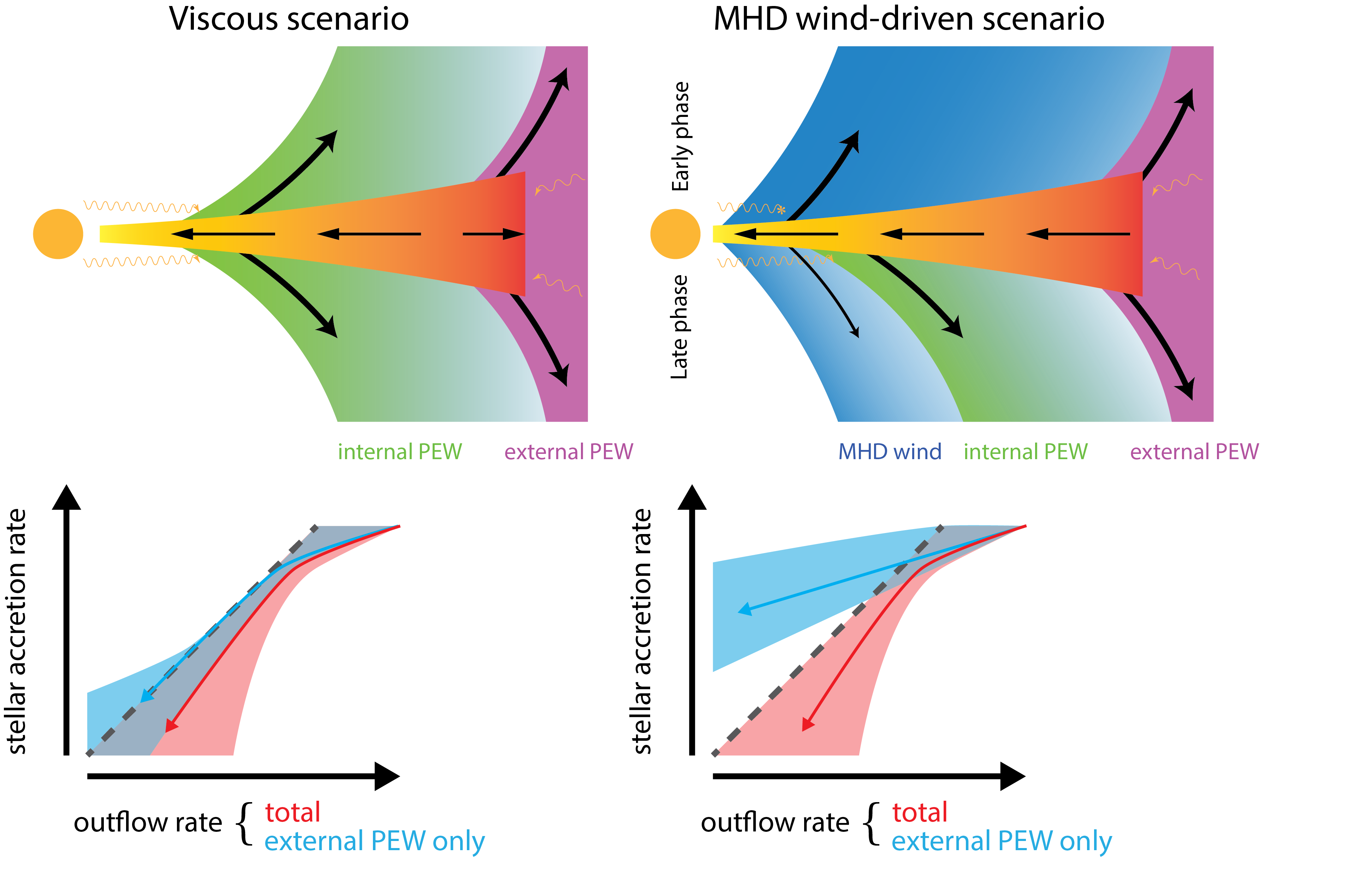}
    \caption{Conceptual illustration of disc evolution scenarios and expected accretion-outflow correlation. Top panels: Conceptual representations of the two evolution scenarios. The protoplanetary disc is shown with arrows indicating the direction of the accretion flow. Outflows emerging from the disc are coloured by different processes i.e. internal photoevaporation (green), external photoevaporation (violet) and MHD wind (blue). Note that in the MHD wind-driven scenario, internal photoevaporation is shielded by the emerging MHD wind in an early phase and only acts in a later phase when the MHD wind has decreased.
    Bottom panels: Expected correlations of the stellar accretion rate versus the outflow rates from all processes (red) or external PEW only (blue), based on theoretical considerations.}
    \label{fig:scenario_overview}
\end{figure*}

While several works have modelled discs in irradiated environments \citep[e.g.][]{Scally2001, Adams2004, Clarke2007, Winter2019b, Concha-Ramirez2019,Qiao2022,coleman_dispersal_2022}, few have attempted to match the properties of discs in both irradiated and non-irradiated regions. It is therefore unclear if the same models applied to reproduce discs in local, low FUV environments also reproduce disc properties in higher FUV regions. If plausible models can be established, it may also be possible to break degeneracies between the viscous and MHD-wind scenarios using irradiated populations. External photoevaporation is most vigorous in the outer disc, and since viscous discs result in outwards spreading, wind mass-loss for evolved, irradiated discs should be sustained at a similar rate to stellar accretion \citep{Clarke2007, Winter2020a}. In an MHD-driven disc, contraction should eventually render external evaporation inefficient. This picture is illustrated in Fig.~\ref{fig:scenario_overview}. The issue angular momentum transport in irradiated disc populations is the focus of this work.

Here we make a direct comparison of the observational signatures of the two evolution scenarios experiencing both internal X-ray and external FUV photoevaporation. We run simulations of synthetic disc populations exposed to low and high ambient FUV fields. For the high FUV field we approximate the distribution of FUV field strengths experienced by stars in intermediate mass star forming regions, with a particular focus on Cygnus OB2. In Sect. \ref{sec:results} we then compare the outcome of the synthetic populations with observables such as disc lifetimes, accretion rates and radii. Further, we test a novel approach of distinguishing between the two evolution scenarios introduced in Sect. \ref{sec:theoretical_understanding}.

% ==================
\section{Characteristics of the two evolution scenarios} \label{sec:theoretical_understanding}
% ==================

%FFFFFFFFFFFFFF
%FFFFFFFFFFFFFF

There are fundamental differences between viscous and MHD wind-driven disc evolution. Here we summarise the characteristics and discuss similarities and differences of the two evolution scenarios.

Global evolution of protoplanetary discs is usually investigated using simple 1D vertically integrated models \citep[e.g.][]{LyndenBell1974,Suzuki2016,Tabone2022a}. For the global disc evolution we here follow the prescription from \cite{Suzuki2016}, \cite{Kunitomo2020} and \cite{Weder2023}. Our model includes both redistribution of angular momentum through turbulent viscosity, angular momentum removal through an MHD wind (MDW) and additional mass loss through thermal winds driven by heating through X-ray irradiation from the central star and FUV irradiation from the surrounding cluster (usually referred to as internal and external photoevaporation (PEW), respectively). The surface density evolution equation takes the form
\begin{equation} \label{eq:disc_evolution}
   \begin{split}
      \frac{\partial \Sigma}{\partial t} =& \frac{1}{r}\frac{\partial}{\partial r}\left[ \frac{3}{r\Omega}\frac{\partial}{\partial r}(r^2\Sigma \overline{\alpha_{r\phi}} c_\mathrm{s}^2) \right] + \frac{1}{r}\frac{\partial}{\partial r}\left[ \frac{2}{\Omega}r \overline{\alpha_{\phi z}}(\rho c_\mathrm{s}^2)_\mathrm{mid} \right] \\
      &- \dot{\Sigma}_\mathrm{MDW} - \dot{\Sigma}_\mathrm{PEW,int} - \dot{\Sigma}_\mathrm{PEW,ext},
  \end{split}
\end{equation}
where $\Sigma$ is the vertically integrated surface density, $r$ corresponds to the distance from the central star, $\Omega$ is the Keplerian frequency and $c_\mathrm{s}$ and $\rho$ correspond to the sound speed and density at the midplane. $\overline{\alpha_{r\phi}}$ is the effective turbulent viscosity and $\overline{\alpha_{\phi z}}$ corresponds to a magnetic stress. Further, there are several sink terms removing mass from the disc, i.e. the magnetic wind $\dot{\Sigma}_\mathrm{MDW}$, internal photoevaporation $\dot{\Sigma}_\mathrm{PEW,int}$ and external photoevaporation $\dot{\Sigma}_\mathrm{PEW,ext}$. Note that the magnetic stress and the magnetic wind are connected as the wind carries away the angular momentum.
The accretion rate through the disc is given by
\begin{equation} \label{eq:disc_accretion}
    \dot{M}_\mathrm{acc}(r) = \frac{6\pi}{r\Omega}\frac{\partial}{\partial r}(r^2\Sigma \overline{\alpha_{r\phi}}c_s^2) + \frac{4\pi}{\Omega}r\overline{\alpha_{\phi z}}(\rho c_s^2)_\mathrm{mid}.
\end{equation}

% ----------------
\subsection{Viscous scenario} \label{sec:viscous_scenario}
% ----------------
In the viscous scenario the disc evolves solely by redistribution of angular momentum and outflows through internal and external photoevaporation. Hence, choosing $\overline{\alpha_{\phi z}}=0$ and $\dot{\Sigma}_\mathrm{MDW}=0$, Equation \ref{eq:disc_evolution} simplifies to
\begin{equation} \label{eq:disc_evolution_viscous}
      \frac{\partial \Sigma}{\partial t} = \frac{1}{r}\frac{\partial}{\partial r}\left[ \frac{3}{r\Omega}\frac{\partial}{\partial r}(r^2\Sigma \overline{\alpha_{r\phi}} c_\mathrm{s}^2) \right] - \dot{\Sigma}_\mathrm{PEW,int} - \dot{\Sigma}_\mathrm{PEW,ext}
\end{equation}
with the accretion rate
\begin{eqnarray} \label{eq:disc_accretion_viscous}
    \dot{M}_\mathrm{acc}(r) &=& \frac{6\pi}{r\Omega}\frac{\partial}{\partial r}(r^2\Sigma \overline{\alpha_{r\phi}}c_s^2) \\
    &=& 3\pi \nu \Sigma + 6\pi r \frac{d}{dr}\left( \nu \Sigma \right), \label{eq:visacc_equi_non_equi}
\end{eqnarray}
where we used the expression for the vertical scale height $H=c_\mathrm{s}/\Omega$ and $\nu=\overline{\alpha_{r\phi}}c_\mathrm{s}H$ \citep{Shakura1973}.

Accretion through the disc is dominated by the equilibrium term (first term in Eq.~\ref{eq:visacc_equi_non_equi}) while the non-equilibrium term (second term in Eq.~\ref{eq:visacc_equi_non_equi}) dominates in the outer disc, where $\Sigma$ drops exponentially, leading to decretion at the same rate due to angular momentum conservation. In the absence of external photoevaporation this leads to disc spreading. However, when including external photoevaporation, disc spreading can be halted or even reversed by the mass removal of the wind. As external photoevaporation is a strong function of radius, decreasing towards smaller radii \citep{haworth_fried_2023}, external photoevaporation will eventually be balanced by decretion leading to an equilibrium state where $\dot{M}_\mathrm{acc,\star}\simeq \dot{M}_\mathrm{PEW,ext}$ \citep{Winter2020a,Hasegawa2022}. $\dot{M}_\mathrm{acc,\star}$ and $\dot{M}_\mathrm{PEW,ext}$ will mutually decrease with time until comparable to the mass-loss-rate from internal photoevaporation $\dot{M}_\mathrm{acc,\star} \simeq \dot{M}_\mathrm{int,PEW}$ shutting down stellar accretion or the disc is dispersed outside-in from the combination of both internal and external photoevaporation. In Figure \ref{fig:scenario_overview} we show the expected behaviour in a schematic plot.

% ------------------------
\subsection{MHD wind-driven scenario} \label{sec:mhd_scenario}
% ------------------------
In the scenario of a purely MHD wind-driven disc evolution (i.e. $\overline{\alpha_{r\phi}}=0$) the evolution equation takes the form
\begin{equation} \label{eq:disc_evolution_mhd}
      \frac{\partial \Sigma}{\partial t} = \frac{1}{r}\frac{\partial}{\partial r}\left[ \frac{2}{\Omega}r \overline{\alpha_{\phi z}}(\rho c_\mathrm{s}^2)_\mathrm{mid} \right] - \dot{\Sigma}_\mathrm{MDW} - \dot{\Sigma}_\mathrm{PEW,int} - \dot{\Sigma}_\mathrm{PEW,ext}
\end{equation}
and the accretion rate is given by
\begin{equation} \label{eq:disc_accretion_mhd}
    \dot{M}_\mathrm{acc}(r) = \frac{4\pi}{\Omega}r\overline{\alpha_{\phi z}}(\rho c_s^2)_\mathrm{mid}.
\end{equation}

The accretion in this scenario is driven by the emerging MHD wind, removing angular momentum from the disc and is often characterised by the magnetic lever arm $\lambda \equiv L/r\Omega$; i.e. the ratio of extracted to initial specific angular momentum \citep{Blandford1982}. From this follows that the stellar accretion rate $\dot{M}_\mathrm{acc,\star}$ and the integrated MHD wind mass-loss-rate $\dot{M}_\mathrm{MDW}$ are related through $\lambda$ (see Fig. 4 \cite{Tabone2022a} and discussion in \citealt{Pascucci2022}). For typical values of the magnetic lever arm $2 < \lambda < 6$ one finds that stellar accretion rate is similar to the wind mass-loss-rate $\dot{M}_\mathrm{acc,\star} \simeq \dot{M}_\mathrm{MDW}$ (see Fig. 4 in \cite{Tabone2022a} and Fig. 9 in \citealt{Weder2023}). Thus, when accounting for all outflows, the stellar accretion rate is expected to correlate with the wind mass-loss-rate until internal photoevaporation engages in the late phase (see Fig. \ref{fig:scenario_overview}). However, in this scenario the disc is accreting through the full extend of the disc and there is no decretion in the outer disc (i.e. no disc spreading). Since the external photoevaporation rate is a strong function of radius it will decrease as the disc shrinks while the disc is still able to accrete at high rates. This fundamental difference when comparing stellar accretion rates with external photoevaporation rates could provide a potential way of identifying the process that drives the evolution of protoplanetary discs, provided that external photoevaporation rates and MHD-wind driven mass loss rates can be disentangled. We test this theoretical prediction in Section \ref{subsec:distinguish_mhd_and_viscous}.

% ==================
\section{Methods} \label{sec:methods}
% ==================
We calculate evolutions of protoplanetary disc populations with varying initial conditions for both viscous and MHD wind-driven evolution models, following closely the approach used in \cite{Weder2023} for MHD wind-driven discs only. The model is summarised in Sect. \ref{subsec:model} and initial conditions of the populations are discussed in Sect. \ref{subsec:init_cond}.

% ------------------
\subsection{Disc evolution model} \label{subsec:model}
% ------------------
The gas disc evolution is modelled using the latest version of the disc module in the Bern Model of Planet Formation and Evolution \citep{burn_toward_2022,Emsenhuber2023,Weder2023}. The evolution is obtained by solving the evolution equation (Eq.~\ref{eq:disc_evolution}). For the vertical structure of the disc, we use the approach of \cite{Nakamoto1994}, including stellar irradiation \citep{Hueso2005} and using the stellar evolution model from \cite{Baraffe2015}.

For the MHD wind-driven disc evolution we are using the model of \cite{Suzuki2016}, where the wind is powered by liberated accretion energy. Based on \cite{Weder2023} where we found that strong winds lead to too short lived discs, we here adopt the weak wind approach where the MHD wind is launched by 10\% of the liberated accretion energy whereas the rest is contributing to disc heating. Furthermore, we chose the magnetic stress $\overline{\alpha_{\phi z}}$ to be constant with time and radius, a simplification that was chosen considering the temporal evolution of the magnetic field remains highly unclear too date. In contrast to \cite{Weder2023} we use a low turbulent viscosity of $\overline{\alpha_{r,\phi}}=10^{-4}$ instead of $\overline{\alpha_{r,\phi}}=5.3\times10^{-5}$ to account for low levels of magneto rotational instability (MRI) and other potential (hydrodynamic) sources of turbulence.

In our model the emerging MHD wind shields the disc from the stellar extreme ultraviolet (EUV) and X-ray radiation in the early phase \citep{Pascucci2022}. As the MHD wind decreases with time, the column density along the disc surface eventually decreases until stellar irradiation can penetrate the wind and internal photoevaporation is switched on leading to a fast dispersal phase \cite[see][for details on the implementation of the shielding mechanism]{Weder2023}.

In contrast to our previous work \citep{Weder2023} we here adopt the internal X-ray photoevaporation model based on \citet{Picogna2019}, \citet{Ercolano2021} and \citet{Picogna2021}, where the latter also include EUV irradiation. The implementation of the model is described in detail in \cite{burn_toward_2022} (see their Sect.~2.2.1). As has been shown in \cite{Emsenhuber2023}, it is difficult to obtain reasonable disc life times when combining high internal and external photoevaporation rates. However, \cite{sellek_photoevaporation_2024} recently showed that the inclusion of cooling from the excitation of O by neutral H leads to a dramatic reduction of the mass-loss rates of previous X-ray photoevaporation models (approximately an order of magnitude for the model from \citealt{Picogna2021}). We therefore divide the obtained mass-loss rates by a factor of 10 which gives the scaling
\begin{equation}
    \dot{M}_\mathrm{PEW,int} \simeq 3.93 \times 10^{-9} \left( \frac{M_\star}{\mathrm{M_\odot}} \right) \mathrm{M_\odot yr^{-1}}.
\end{equation}
We would like to point out, that although we here consider X-ray as the driver of internal photoevaporation, it has been shown that FUV from the host star is also capable of driving winds at similar mass-loss rates \citep{Gorti2015,Nakatani2018a,Nakatani2018b}. Different models producing similar mass-loss rates and profiles are not expected to qualitatively influence our results. We would further like to add that we have checked our results using lower mass-loss rates from EUV photoevaporation and found no qualitative changes.

Further, we are now using the new FRIED grid v2 \citep{haworth_fried_2023} to obtain mass-loss-rates for external photoevaporation for varying ambient FUV field strength $\mathcal{F}_\mathrm{FUV}$, stellar mass $M_\star$, disc mass $M_\mathrm{disc}$ and size $r_\mathrm{out}$. The new version provides mass-loss-rates for FUV field strengths down to $\mathrm{1\,G_0}$ and additional grids for different polycyclic aromatic hydrocarbon abundances ($f_\mathrm{PAH}$) with and without grain growth. We here only consider the case including grain growth. The obtained mass-loss-rates are removed from the outermost 10\% of the radial extent of the disc. Details about the implementation can be found in \cite{Weder2023}.

% ------------------
\subsection{Initial conditions} \label{subsec:init_cond}
% ------------------

The initial conditions for the synthetic populations are chosen to reflect conditions found in young star forming regions. We follow closely \cite{Weder2023} with changes regarding the stellar mass distribution and we use the approach in \cite{winter_external_2019} to calculate an FUV flux distribution in the massive Cygnus OB2 star cluster (see Sec.~\ref{subsubsec:modelling_star_formation_FUV_field_CygnOB2}). The obtained cumulative distributions for the initial conditions are shown in Figure \ref{fig:init_cond}.

% - - - - - - - - - - 
\subsubsection{Modelling star formation and the FUV field in Cygnus OB2} \label{subsubsec:modelling_star_formation_FUV_field_CygnOB2}
% - - - - - - - - - - 
We aim to reproduce a realistic distribution of FUV flux histories experienced by the stellar population in typical star forming regions. This requires a model for the star formation rate, population of massive stars, interstellar extinction, and spatial distribution of young stellar objects (YSO). Although we aim to make our results as general as possible, we choose to base our model on the observed properties of Cygnus OB2. We choose Cygnus OB2 because:
\begin{enumerate}
    \item The stellar population is relatively well studied \citep[e.g.][]{wright_massive_2015}.
    \item Both the stellar densities and range of FUV fluxes experienced by the YSOs in that region is fairly typical ($\sim 10^{2}-10^5\,\mathrm{G_0}$) and similar to those in a number of other local star forming regions \citep[e.g.][]{winter_protoplanetary_2018, anania_novel_2025}.
    \item The overall number of stars is much higher, such that statistical averaging in a number of FUV bins is possible. 
    \item Cygnus OB2 is one of the few regions in which there is a clear gradient in fraction of surviving discs (measured by infrared excess) is observed as a function of external FUV field \citep{guarcello_photoevaporation_2016, Guarcello2023}. 
\end{enumerate}In the following, we therefore discuss the assumed properties of Cygnus OB2, but it should be understood that our results should broadly generalise to any star forming region with a similar distribution of FUV fluxes experienced by the young stellar population.

The total stellar mass of the Cygnus OB2 association is $\sim 1.6\times 10^{4}\,\mathrm{M_\odot}$ \citep{wright_massive_2015}. The majority of stars formed between $3-5\,\mathrm{Myr}$ ago \citep{wright_massive_2010}, however with stellar ages spanning a large range from $\sim 2\,\mathrm{Myr}$ \citep{hanson_study_2003} to $\sim7\,\mathrm{Myr}$ \citep{drew_early-stars_2008}. 

\begin{figure}
    \centering
    \includegraphics[width=\linewidth]{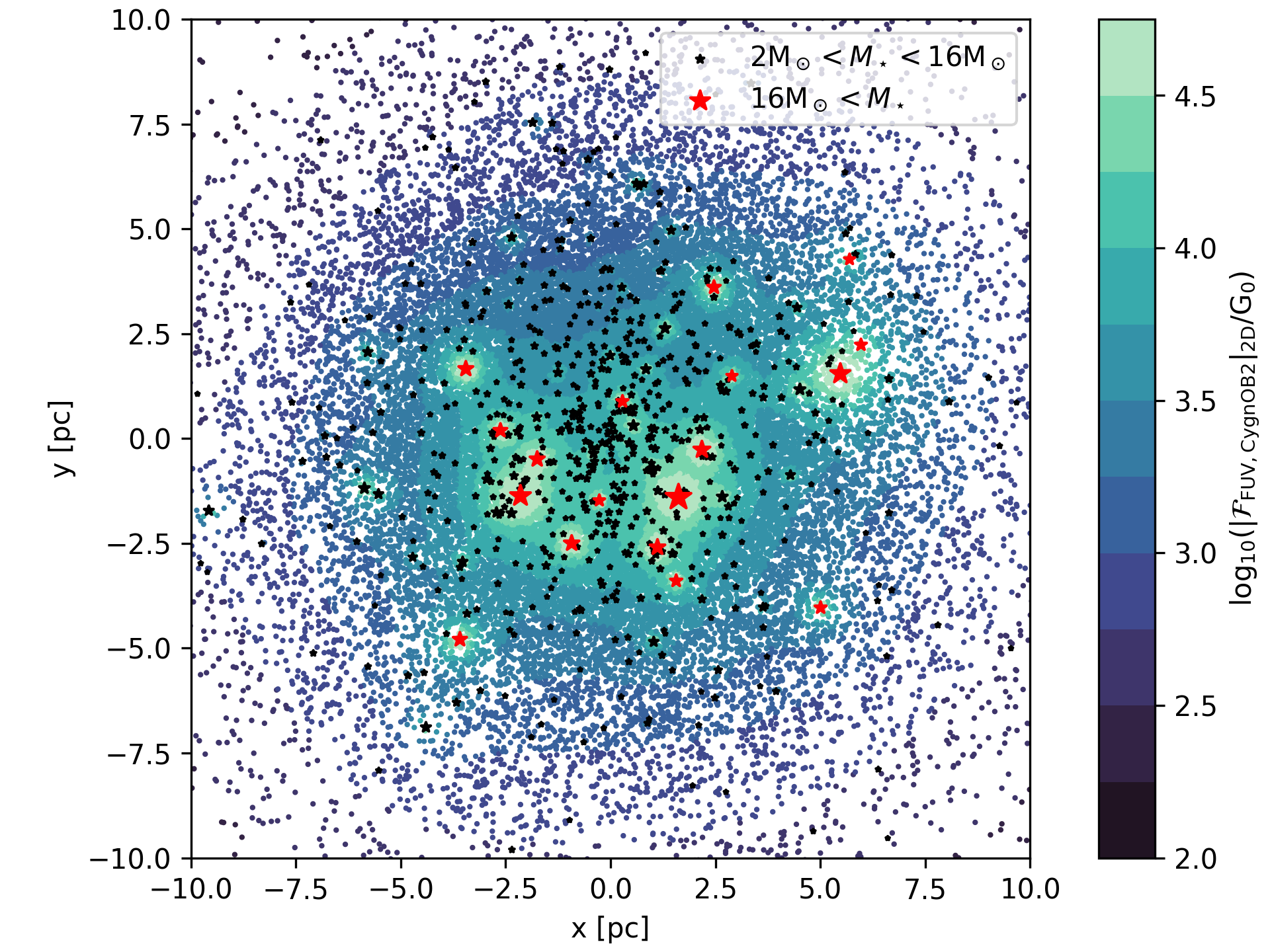}
    \caption{View of FUV field in the synthetic cluster after the formation phase ($t=\mathrm{2\,Myr}$). Stars are coloured by the projected FUV field $|\mathcal{F}_\mathrm{FUV,CygnOB2}|_\mathrm{2D}$. Stars contributing to the local FUV field are highlighted by black ($\mathrm{2\,M_\odot}<M_\star<\mathrm{16\,M_\odot}$) or red ($16\,\mathrm{M_\odot}<M_\star$) markers. The size of the markers scale with stellar mass.}
    \label{fig:cluster}
\end{figure}

To estimate the evolution of the external FUV flux experienced by the YSO population, we adopt the following recipe. We assume an initial molecular cloud with a gas mass of $M_{g}(t=0) \simeq 1.5 \times 10^{6}\,\mathrm{M_\odot}$. Assuming a star formation efficiency of $\mathrm{SFE}=0.01$ this will lead to a cluster of $\simeq 1.5 \times 10^{4}\,\mathrm{M_\odot}$, which is similar to Cygnus OB2. We assume that the star formation rate was uniform during a period of $\mathrm{2\,Myr}$, assigning a formation time to each star. Following \cite{winter_external_2019} we sample the initial mass function (IMF) \cite{kroupa_variation_2001} and assign coordinates $\vec{r}_i=(x,y,z)$ according to an EFF profile (i.e. Elson, Fall and Friedman profile, \cite{elson_structure_1987})
\begin{equation}
    \rho_\star(r) = \rho_0 \left( 1 + \frac{r^2}{a^2_\mathrm{stars}} \right)^{-\frac{\gamma + 1}{2}}
\end{equation}
with $a_\mathrm{stars}=7.5\mathrm{pc}$ and $\gamma=5.8$ \citep{wright_massive_2015}. The remaining mass of the molecular cloud is then given at each time by 
\begin{equation}
        M_\mathrm{g}(t)= M_\mathrm{g}(t=0) - \frac{\sum_{t_i<t}M_{\star,i}}{\mathrm{SFE}},
\end{equation}
with $t_i$ being the time when the star is formed. The density profile of the gas in the molecular cloud is simply given by
\begin{equation}
    \rho_\mathrm{g}(r,t)=\frac{M_\mathrm{g}(t)}{\frac{4}{3} \pi a_\mathrm{stars}^3} \cdot \left( 1+\frac{r^2}{a_\mathrm{stars}^2} \right)^{-\frac{\gamma+1}{2}}.
\end{equation}
Assuming that all stars with masses $M_\star>2\,\mathrm{M_\odot}$ contribute to the FUV flux, the local FUV field of each star is then calculated by summing up the contributions of all stars present at that time $t$, including the shielding by the remnant gas. FUV luminosities $L_\mathrm{FUV}$ for the massive stars are calculated based on the MIST isochrones \citep{Paxton2013, Choi2016, Dotter2016} and PHOENIX atmosphere models \citep{Husser2013}. The interstellar shielding is calculated by integrating the gas density along the line of sight between two stars $i$ and $j$, obtaining a surface density
\begin{equation}
    \Sigma_{i,j}^{\mathrm{ext}}=\left| \int_{\mathcal{C}(r_i,r_j)}\rho_{g}(t,r)d\vec{r} \right|,
\end{equation}
with $r_i$ and $r_j$ being the positions of stars $i$ and $j$ and the path $\mathcal{C}$ the straight line in between. Using the FUV to visual extinction $A_\mathrm{FUV}/A_\mathrm{V}=2.7$ \citep{cardelli_relationship_1989} and the column density of hydrogen required for $1\,\mathrm{mag}$ of extinction $N_\mathrm{H}/A_\mathrm{V}=1.8\times 10^{21}\mathrm{cm^{-2}mag^{-1}}$ \citep{predehl_x-raying_1995}, this can be converted into an extinction
\begin{equation}
    E_{i,j}(t)=10^{-0.4 \times 2.7 \frac{\Sigma_\mathrm{i,j}^\mathrm{ext}/(\mu_\mathrm{H} \cdot m_\mathrm{H})}{1.8\cdot 10^{21}\,\mathrm{{cm^{-2}}}}},
\end{equation}
where $\Sigma_\mathrm{i,j}^\mathrm{ext}/(\mu_\mathrm{H} \cdot m_\mathrm{H})$ corresponds to the hydrogen column density between stars $i$ and $j$, with the mean molecular weight per hydrogen atom $\mu_\mathrm{H}=1.4$ and hydrogen atom mass $m_\mathrm{H}$.
By summing up all the contributions, we obtain a self-consistent, time dependent FUV field for each star
\begin{equation} \label{eq:3D_FUV_field}
    \mathcal{F}_{\mathrm{FUV},i}(t) = \sum_{j \neq i} \frac{L_{\mathrm{FUV},j} \times E_{i,j}(t)}{4\pi |\vec{r}_j-\vec{r}_i|^2} \qquad \mathrm{t_i<t_j \,and\,}M_j>2\mathrm{M_\odot}.
\end{equation}
We label this spatially resolved FUV field of Cygnus OB2 $\mathcal{F}_\mathrm{FUV,CygnOB2}$. However, observations lack precise 3D locations of the stars within the cluster \citep[see discussion by][for example]{anania_novel_2025}. Therefore 2D projected maps are used \cite[e.g.][]{guarcello_photoevaporation_2016}. The projected 2D FUV field of the cluster after the formation phase is obtained using Eq. \ref{eq:3D_FUV_field}, however replacing the position vectors by the 2D projections $\vec{r}_i=(x,y,0)$. In Figure \ref{fig:cluster} we show the synthetic cluster after formation ($t>2\,\mathrm{Myr}$) with stars being coloured by the 2D projected local FUV field $|\mathcal{F}_\mathrm{FUV,CygnOB2}|_\mathrm{2D}$. It nicely illustrates how the FUV field is dominated by the most massive stars.

Unlike \citet{winter_external_2019}, who also modelled disc evolution in Cygnus OB2, we do not attempt to model the dynamical evolution or sub-structure in Cygnus OB2. This is partly due to the complex picture, which is challenging to model self-consistently. Globally, Cygnus OB2 has a very large velocity dispersion \citep[$\sim 17$~km~s$^{-1}$][]{Wright2016}, implying a highly supervirial and expanding population. However, \citet{winter_external_2019} found that the nominal age and total stellar mass results in expansion too rapid to produce the current spatial distribution of stars. The solution of those authors was to add an extremely massive (spherically symmetric) background potential, to globally contain the stars until very recently. This is plausible, but in a sub-structured region containing many massive stars, we expect gas to be expelled rapidly \citep{Qiao2022, Wilhelm2023}. This would particularly influence the gas in the most irradiated regions, and therefore reduce the overall extinction due to residual gas with respect to our spherically symmetric model. A complete dynamical model of the region is therefore highly challenging. Our focus here is more general than just the population of Cygnus OB2, and our scope is more related to the evolution of the discs themselves. We therefore ignore dynamics and structure, but expect that a more elaborate model including them would result in larger (historic) FUV fluxes for the observed YSOs. How this might change our results is discussed in Sect.~\ref{subsec:cygnusOB2_compar}. 

%FFFFFFFFFFFFFF
%FFFFFFFFFFFFFF

% - - - - - - - - - - 
\subsubsection{Initial properties of the synthetic populations} \label{subsubec:initial_disc_properties}
% - - - - - - - - - - 
The synthetic population is constructed by selecting stars from the population and assign them a disc of mass $M_\mathrm{disc}$ and size $R_\mathrm{char}$ with an inner edge $R_\mathrm{in}$.

\paragraph{Star selection}
Surveys of stellar populations in star forming regions often suffer from incompleteness and small number statistics \cite[see discussion in e.g.][]{Testi2022a}. While some of the low mass regions (i.e. Lupus, Chamaeleon) are close to complete down to $\sim0.1\,\mathrm{M_\odot}$ \citep{luhman_gaia_2020}, surveys of massive star forming regions are often incomplete, especially towards lower mass, due to their larger distances. The sample used in \cite{guarcello_photoevaporation_2016} shows decay of number of stars below $0.3\,\mathrm{M}_\odot$ which indicates the incompleteness below that mass (M. Guarcello,  personal communication, January, 2024). Modelling protoplanetary discs around low mass stars poses a challenge as disc lifetimes start to decrease rapidly with stellar mass \cite[see Figs. 2 and 3 in][]{Emsenhuber2023}. This is in contrast to observations that show an increasing disc fraction with lower stellar mass \citep{luhman_census_2022,Pfalzner_discFraction_2022}. In disc population syntheses, the evolution of the disc fraction is very sensitive to the lower mass stars as they are much more abundant. However, the disc lifetime is not expected to increase strongly with decreasing stellar mass $\lesssim0.3\,\mathrm{M_\odot}$ and therefore choosing a lower limit of $0.3\,\mathrm{M_\odot}$ is a reasonable choice even tough the completeness between low and high mass clusters vary. We therefore limit the range of stellar masses in the population to $0.3\,\mathrm{M_\odot}<M_\star<2\,\mathrm{M_\odot}$. We further assume that 80\% of stars host a protoplanetary disc in order to account for close in binaries that may quickly remove inner disc material \cite{Kraus2012}.

\paragraph{Disc properties}
Following \cite{Weder2023} we use a log-normal fit to dust disc masses of Class 1 protoplanetary discs representative for stellar masses of $\sim 0.3\mathrm{M_\odot}$ \cite[i.e. $\log_{10}(\mu/\mathrm{M_\oplus})=2.03$ and $\sigma=0.35\,\mathrm{dex}$,][]{Tychoniec2018}, applying a linear scaling with stellar mass. The characteristic radius is then calculated by using the relation $R_\mathrm{char}=70\,\mathrm{au} \cdot \left[ M_\mathrm{dust} / 100 \mathrm{M_\oplus} \right]^{0.25}$ with a $0.1\,\mathrm{dex}$ spread, similar to what has been found in \cite{Tobin2020}. To obtain estimations for the gas mass, we use a dust-to-gas ratio distribution that is based on a normal distribution from observed metallicity \cite[i.e. $\mu=-0.02$ and $\sigma=0.22$][]{Santos2005} and use the relation $f_\mathrm{D/G}=f_\mathrm{D/G,\odot}10^{[\mathrm{Fe/H}]}$. Assuming the inner radius of the disc corresponds to the corotation radius, we use a log-normal fit to observed rotation rates from \cite{Venuti2017}.
The initial surface density profile is then set as
\begin{equation} \label{eq:init_surface_density}
   \Sigma_\mathrm{init}(r) = \Sigma_\mathrm{0,5.2au} \left( \frac{r}{5.2\,\mathrm{au}} \right)^{-\beta} e^{-\left( \frac{r}{R_\mathrm{char}} \right)^{2-\beta}} \left( 1-\sqrt{\frac{R_\mathrm{in}}{r}} \right),
\end{equation}
with $\beta=0.9$ and $\Sigma_\mathrm{0,5.2au}$ being the appropriate value to obtain the assigned disc mass. For more details regarding the initial conditions of the discs, we refer to Section 2.2 in \cite{Weder2023}.

\begin{figure}[h]
    \includegraphics[width=\linewidth]{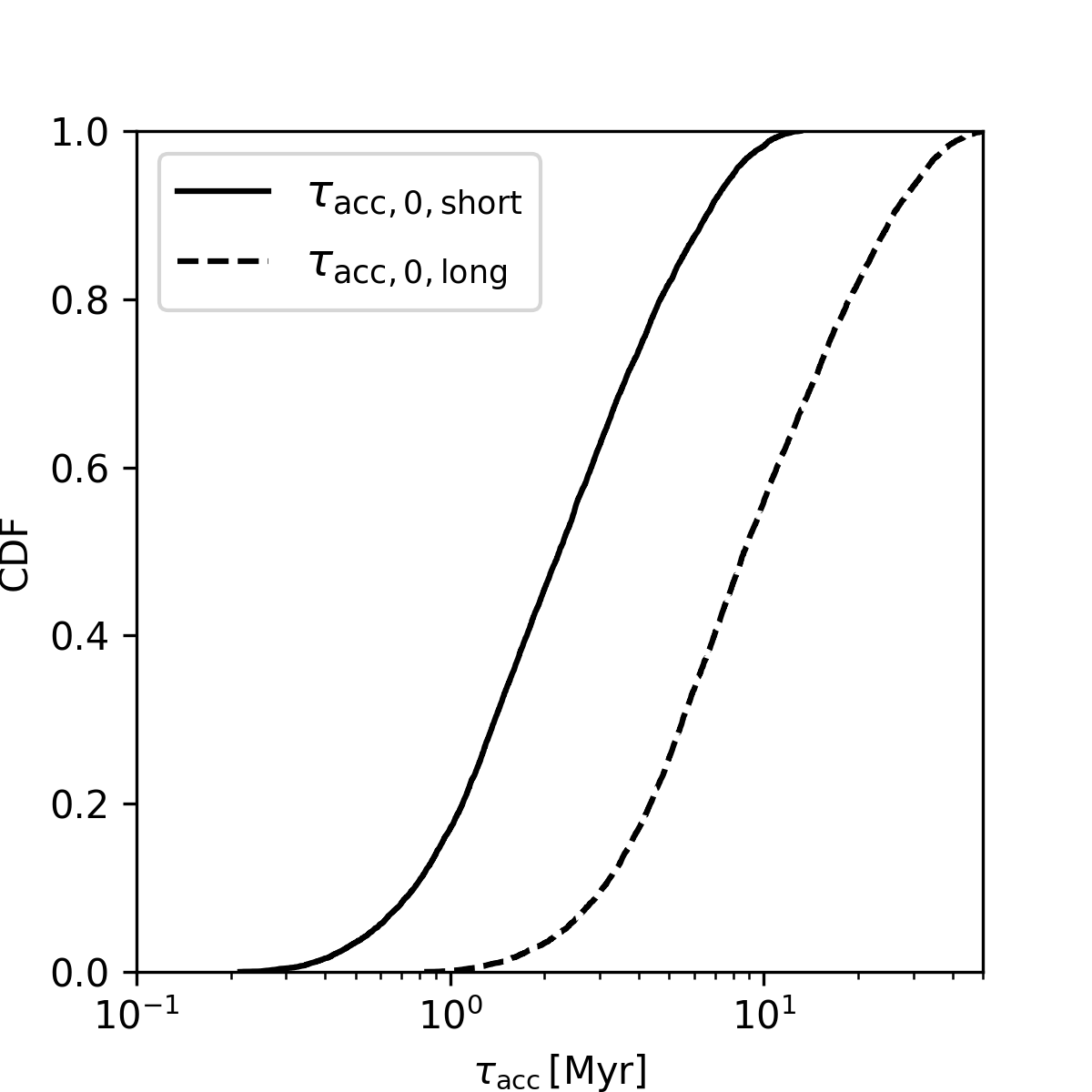}
    \centering
    \caption{Considered initial accretion timescale distributions $\tau_\mathrm{acc,0,short}$ and $\tau_\mathrm{acc,0,long}$.}
    \label{fig:acc_timescale}
\end{figure}

\paragraph{Angular momentum transport}
Disc evolution itself is governed by $\overline{\alpha_{r\phi}}$ and $\overline{\alpha_{\phi z}}$ for the viscous and MHD wind-driven scenario respectively. We use the initial accretion timescale $\tau_\mathrm{acc,0}= M_\mathrm{disc} / \dot{M}_\mathrm{acc}$ to define sets of $\overline{\alpha_{r\phi}}$ and $\overline{\alpha_{\phi z}}$ that give similar disc lifetimes and accretion rates. Using the analytical expressions for the accretion rates (Eq.~\ref{eq:disc_accretion_viscous} and \ref{eq:disc_accretion_mhd}), we can derive expressions for both $\overline{\alpha_{r\phi}}$ and $\overline{\alpha_{\phi z}}$ that feature $\tau_\mathrm{acc,0}$
\begin{eqnarray}
    \overline{\alpha_{r\phi}} &=& \frac{M_\mathrm{disc}}{3\pi r^2 h^2 \Omega \Sigma \tau_\mathrm{acc,0}} \label{eq:eq:alpha_acc_timescale_vis} \\
    \overline{\alpha_{\phi z}} &=& \frac{M_\mathrm{disc}}{2\sqrt{2\pi} r^2 h \Omega \Sigma \tau_\mathrm{acc,0}} \cdot f_c. \label{eq:alpha_acc_timescale_mhd}
\end{eqnarray}
We introduced a correction factor $f_c$, which is a free parameter that is chosen such that the lifetimes agree better for a given $\tau_\mathrm{acc,0}$. Note that we only consider the equilibrium term for the diffusive part as the non equilibrium part is negligible through large parts of the disc. Expressions \ref{eq:eq:alpha_acc_timescale_vis} and \ref{eq:alpha_acc_timescale_mhd} are evaluated at the characteristic radius (i.e. where the bulk of the mass lies) and $f_c=0.8$ was found empirically to be a good fit. Observational constraints reveal a broad range of accretion timescales among young stars and between different star forming regions (e.g. Fig.~12 in \citealt{almendros-abad_evolution_2024}). We adopt two distributions of the initial accretion timescales that are uniform in log and scaled linear with stellar mass: $\log(\tau_\mathrm{acc,0,short} \times M_\star/\mathrm{M_\odot})=\mathcal{U}(-0.4,+0.6)$ and $\log(\tau_\mathrm{acc,0,long} \times M_\star/\mathrm{M_\odot})=\mathcal{U}(0.2,1.2)$, see Fig.~\ref{fig:acc_timescale}. Note that the scaling of $\tau_\mathrm{acc}$ with stellar mass gives us the relation $\dot{M}_\mathrm{acc} \propto M_\star^2$ which is in line with observational constraints with slopes of $\sim 1.6 - 2$ \cite[see review by][]{Manara2022}. We point out that the initial accretion timescales adopted here are empirically found to be able to reproduce observed disc fractions and evolve with time. Choosing shorter $\tau_\mathrm{acc,0}$ leads to shorter lived discs, which is in contrast to observed disc lifetimes (see Sect.~\ref{subsec:disc_frac_evo}). The resulting distributions for $\overline{\alpha_{r\phi}}$ and $\overline{\alpha_{\phi z}}$ are shown in the bottom right panel of Fig. \ref{fig:init_cond} in Appendix \ref{app:initial_conditions}.
%FFFFFFFFFFFFFF
%FFFFFFFFFFFFFF

% - - - - - - - - - - 
\subsubsection{Stellar X-ray luminosity} \label{subsubsec:xray_lumi}
% - - - - - - - - - - 
For the stellar X-ray luminosity we follow \cite{Emsenhuber2023} where we used the results from the XMM-Newton Extended Survey of Taurus (XEST) from \cite{Guedel2007}, who found a stellar mass dependence of $L_\mathrm{X}\propto M_\star^{1.45\pm0.12}$. We adopt a log-normal distribution for the X-ray luminosity with $\log_\mathrm{10}(\mu/\mathrm{erg\,s^{-1}})=30.31 + 1.52 \times \log_\mathrm{10}(M_\star/\mathrm{M_\odot})$ and a $\sigma=0.54\,\mathrm{dex}$ spread. The resulting $L_\mathrm{X}$ distribution is shown in panel c) of Fig.~\ref{fig:init_cond} in Appendix \ref{app:initial_conditions}.\\

% - - - - - - - - - - 
\subsubsection{Low FUV environment} \label{subsubsec:init_cond_low_FUV_environment}
% - - - - - - - - - - 
To contextualise our models, we aim to explore whether they reproduce the lifetimes of discs in the local, low mass and FUV regions (e.g. Lupus, Chamaeleon I). We therefore run additional sets of simulations with a low ambient FUV field of $\mathcal{F}_\mathrm{FUV,10G_0}=\mathrm{10\,G_0}$. In this case, since we do not need to evolve the FUV field, we assume all stars are formed instantaneously. This will influence the disc fractions purely due to the lack of age spread, however these differences are negligible when averaged over stars.

%==============
\section{Results} \label{sec:results}
%==============
We ran simulations using the initial conditions described in Section \ref{subsec:init_cond}. We additionally varied the polycyclic aromatic hydrocarbon abundances $f_\mathrm{PAH}=\{1,0.5,0.1\}$. Since photoelectric heating by PAHs is probably a dominant heating source for the externally driven wind, $f_\mathrm{PAH}$ is an important but uncertain quantity in determining the externally driven mass-loss rate.

Populations including the strong FUV field $\mathcal{F}_\mathrm{FUV,CygnOB2}$ consist of $10\,000$ systems in order to have good statistics in the low FUV bins (see Section \ref{subsec:cygnusOB2_compar}). Populations with a weak FUV field $\mathcal{F}_\mathrm{FUV,10G_0}$ consist of $2\,000$ simulations each.

%FFFFFFFFFFFFFF
\begin{figure*}[ht!]
    \centering
    \includegraphics[width=\linewidth]{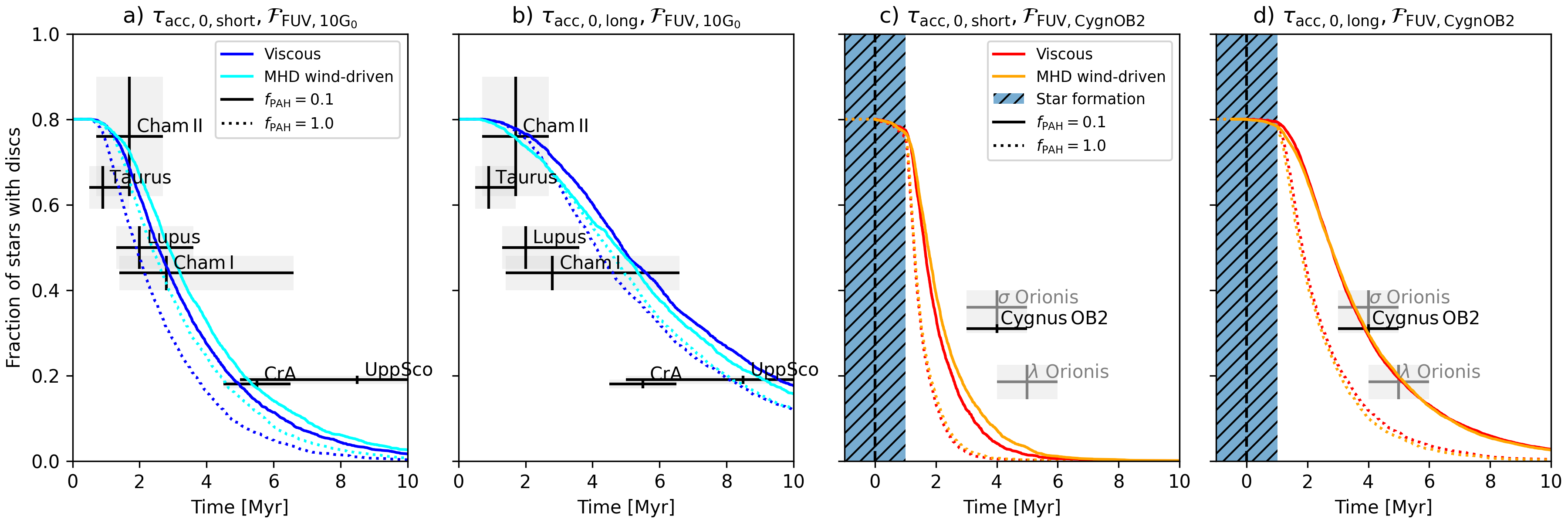}
    \caption{Time evolution of the disc fraction for different evolution scenarios, FUV field strengths $\mathcal{F}_\mathrm{FUV,10G_0}$ and $\mathcal{F}_\mathrm{FUV,CygnOB2}$, varying $\tau_\mathrm{acc}$ and $f_\mathrm{PAH}$. Note that in the strongly irradiated cases, a duration of star formation of 2 Myr is accounted for while the low irradiated cases stars are set at $t=0$. Panels a) and b) show results from simulations with $\mathcal{F}_\mathrm{FUV,10G_0}$ for $\tau_\mathrm{acc,0,short}$ and $\tau_\mathrm{acc,0,long}$, alongside with a selected sample of observed disc fractions for low mass star forming regions. Panels c) and d) show the results of the strongly irradiated cases with $\mathcal{F}_\mathrm{FUV,CygnOB2}$ for $\tau_\mathrm{acc,0,short}$ and $\tau_\mathrm{acc,0,long}$, alongside with observed disc fraction in Cygnus OB2 and two additional high mass star forming regions. See Appendix \ref{app:observed_disc_fractions} for a Table \ref{tab:disc_fractions} and discussion of the observational constraints.}
    \label{fig:disc_fraction_evo}
\end{figure*}
%FFFFFFFFFFFFFF

We start by assessing the disc lifetimes of the different populations by comparing the time evolution of the disc fraction with observations in Section \ref{subsec:disc_frac_evo}. We further show the $M_\mathrm{disc}-\dot{M}_\mathrm{acc}$ relation for a subset with $f_\mathrm{PAH}=0.1$ in Section \ref{subsec:mdisc_macc}. From this, we identify accretion timescales $\tau_\mathrm{acc,0}$ and PAH abundances $f_\mathrm{PAH}$ that yield disc lifetimes and stellar accretion rates that are in line with observations. This results in a set of four populations (viscous and MHD wind driven scenarios with strong and weak FUV field), for which we investigate the spatial variation in disc fraction of our synthetic massive cluster in Section \ref{subsec:cygnusOB2_compar}, compare the evolution of disc radii in Section \ref{subsec:disc_radii} and test our prediction regarding different evolution of accretion rate vs. wind-mass-loss rates in Section \ref{subsec:distinguish_mhd_and_viscous}.

% -------------
\subsection{Disc lifetimes}\label{subsec:disc_frac_evo}
% -------------

The lifetime of protoplanetary discs is one of the most important constraints as it sets the duration during which planets can be formed. As the determination of the age and stage of individual systems is not feasible and depends on many aspects (e.g. presence of OB type stars, cluster dynamics), disc lifetime is often regarded as an evolution of the fraction of stars remaining with a disc (typically inferred by the presence/absence of an infrared excess) in clusters of different age \citep[e.g.][]{HaischJr.2001}. As already demonstrated in \cite{Weder2023}, external photoevaporation can have a significant effect on the disc lifetime. Following \cite{Weder2023}, the disc is considered dispersed if the optical depth $\tau$ drops below unity in regions where the midplane temperature is $T_\mathrm{mid}>\mathrm{300K}$ \citep{Kimura2016}.

In Figure \ref{fig:disc_fraction_evo} we show the evolution of the disc fraction for simulations with $f_\mathrm{PAH}=\{0.1,1\}$. Simulations with intermediate $f_\mathrm{PAH}=0.5$ have lifetimes in between the aforementioned values. We show the results for $\mathcal{F}_\mathrm{FUV,10G_0}$ with either $\tau_\mathrm{acc,0,short}$ or $\tau_\mathrm{acc,0,long}$ and $\mathcal{F}_\mathrm{FUV,CygnOB2}$ with $\tau_\mathrm{acc,0,short}$ or $\tau_\mathrm{acc,0,long}$ in separate figures for better readability.

Although the two disc evolution process are fundamentally different, the choice of similar $\tau_\mathrm{acc}$ (with 25\% longer accretion timescales for the MHD wind-driven scenario due to the chosen correction factor $f_c=0.8$) leads to overall excellent agreement between both evolution scenarios.

The panels of Figure \ref{fig:disc_fraction_evo} can be summarised as follows:
\begin{enumerate}[a)]
    \item shows the results of simulations in $\mathcal{F}_\mathrm{FUV,10G_0}$ with $\tau_\mathrm{acc,0,short}$. The simulations show good agreement with the observational trend of disc dispersion in low irradiated regions (see Appendix \ref{app:observed_disc_fractions}). Varying the $\mathrm{PAH}$ abundance $f_\mathrm{PAH}$ shows only little effect. High $\mathrm{PAH}$ abundances lead to higher external photoevaporation rates and shorter lived discs, whereas lower $\mathrm{PAH}$ abundances lead to lower external photoevaporation and thus to longer lived discs.
    
    \item shows the results of simulations in $\mathcal{F}_\mathrm{FUV,10G_0}$ with $\tau_\mathrm{acc,0,long}$. The simulations with low PAH abundances show slower evolution of the disc fractions. Overall we find reasonable agreement with observed disc fractions.
    
    \item shows the strongly irradiated cases with $\mathcal{F}_\mathrm{FUV,CygnOB2}$ and $\tau_\mathrm{acc,0,short}$. The disc fractions drop rapidly after the formation phase and reach $\lesssim 0.4$ after $\sim2\,\mathrm{Myr}$, regardless of the assumed $f_\mathrm{PAH}$. This is lower than the observed disc fractions in star forming regions with similarly high FUV flux distributions.
    
    \item shows the simulations with $\mathcal{F}_\mathrm{FUV,CygnOB2}$ and $\tau_\mathrm{acc,0,long}$. A low PAH abundance of $f_\mathrm{PAH}=0.1$ can reproduce the disc fractions in highly irradiated star forming regions. Higher PAH abundances result again in too short lived discs.
\end{enumerate}

Our simulations show that the accretion timescale has significant impact on the disc lifetimes in both low and high FUV field environments, while the PAH abundance shows a larger influence for highly irradiated regions since here the lifetime is largely set by the external photoevaporation. While for $\mathcal{F}_\mathrm{FUV,10G_0}$ the disc fraction match observations, regardless of the choice of $\tau_\mathrm{acc,0}$ and $f_\mathrm{PAH}$, in the case of $\mathcal{F}_\mathrm{FUV,CygnOB2}$ we need both low PAH abundance and long accretion timescales in order to reproduce observed disc fractions in highly irradiated regions. This might then imply long accretion time-scales for discs generally. However, observational constraints on low UV regions show short accretion timescales of ${\sim} 1\,\mathrm{Myr}$ (see Figure~7 \citealt{Manara2022}), which would lead to too short lived discs. Therefore these results point towards a fundamental problem in reproducing both irradiated and non-irradiated discs with a single set of models (see also Section~\ref{subsec:mdisc_macc}).

% -------------
\subsection{$M_\mathrm{disc} - \dot{M}_\mathrm{acc,\star}$ relation} \label{subsec:mdisc_macc}
% -------------
%FFFFFFFFFFFFFF
\begin{figure*}[htbp]
    \centering
    \begin{subfigure}{0.45\textwidth}
      \centering
      \includegraphics[width=\textwidth]{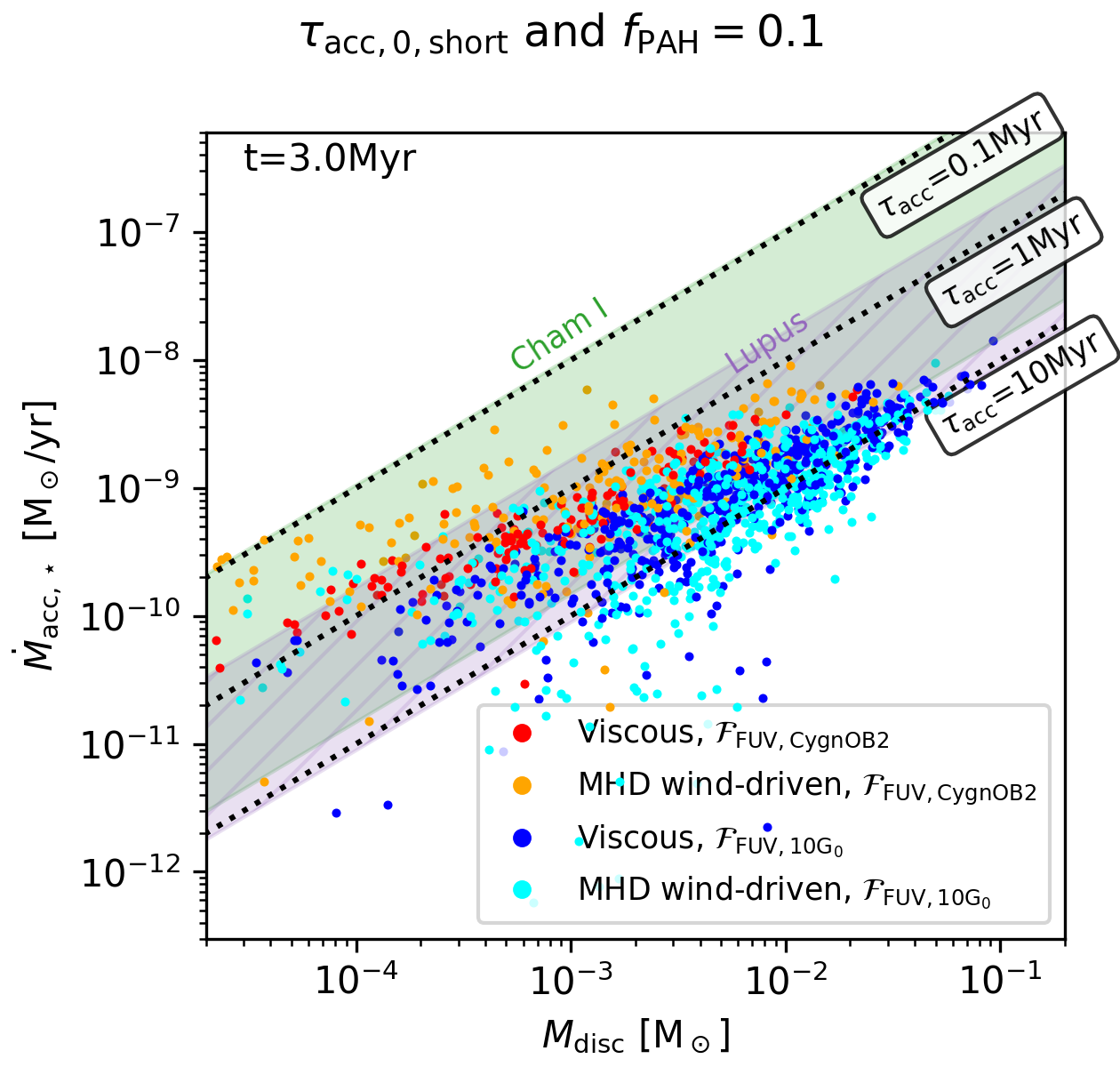}
    \end{subfigure}
    \hfill
    \begin{subfigure}{0.45\textwidth}
      \centering
      \includegraphics[width=\textwidth]{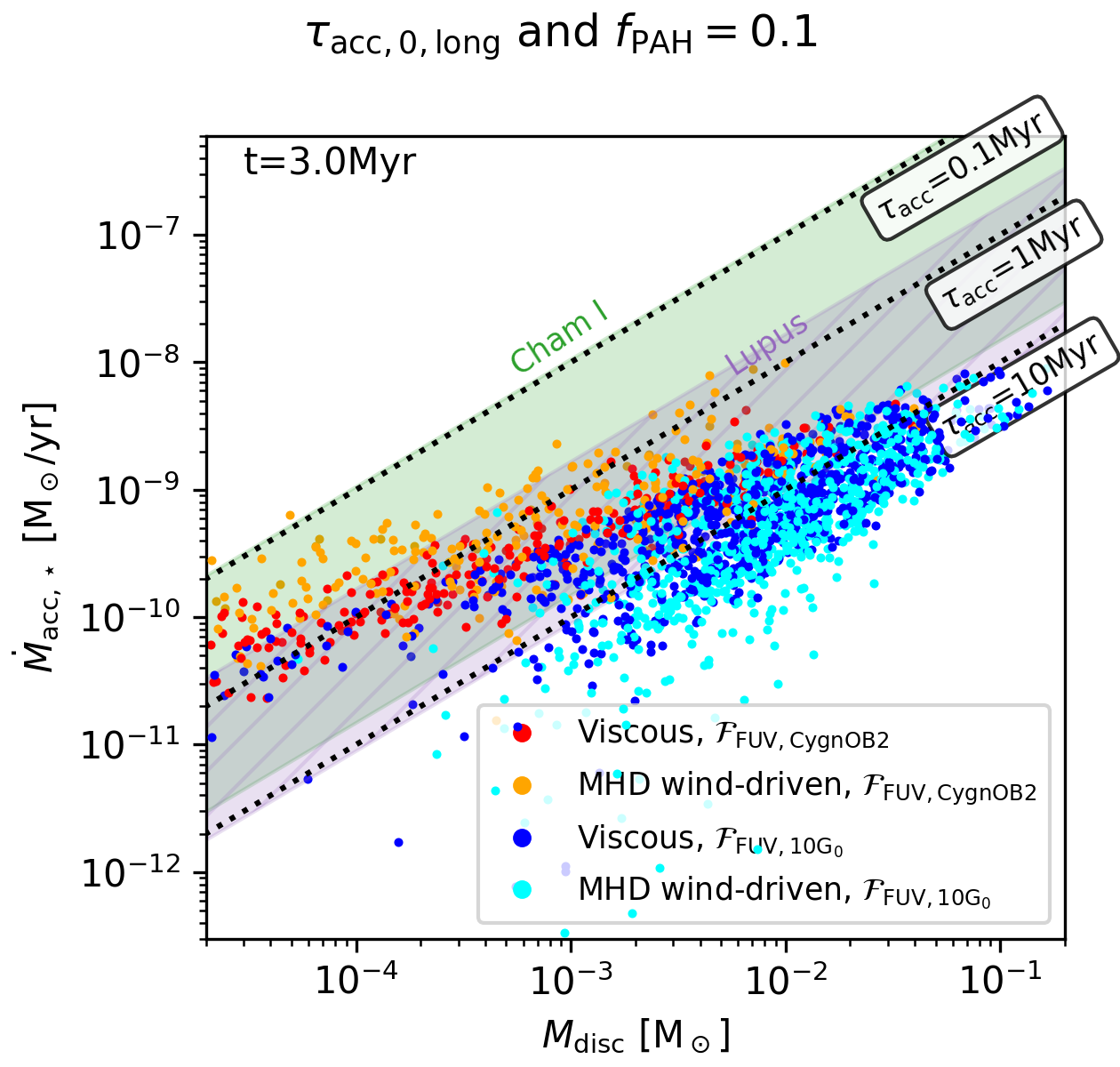}
    \end{subfigure}
    \caption{Stellar accretion rate $\dot{M}_\mathrm{acc,\star}$ versus disc masses $M_\mathrm{disc}$ for simulations with $f_\mathrm{PAH}=0.1$ and $\tau_\mathrm{acc,0,short}$ (left panel) and $\tau_\mathrm{acc,0,long}$ (right panel). A subset of $\mathrm{1000}$ simulations is shown at $\mathrm{3\,Myr}$. Lines of constant $M_\mathrm{disc}/\dot{M}_\mathrm{acc,\star}$ (i.e. $\tau_\mathrm{acc}$) are shown for 0.1, 1 and 10 Myr (dotted lines). The coloured areas show the central 60\% of the $\tau_\mathrm{acc}$ distribution for close-in low mass star forming regions Lupus and Chamaeleon I, obtained from \citealt{Manara2022}. The large difference between our low FUV simulations and observations from Chamaeleon I and Lupus point towards strongly variable, time-dependent and heterogeneous angular momentum transport between different regions.}
    \label{fig:mdisc_vs_macc}
\end{figure*}
%FFFFFFFFFFFFFF
Another important observational constraint is the relation of stellar accretion rate and disc mass \citep{ManaraMordasini2019,Manara2022}. In Figure~\ref{fig:mdisc_vs_macc} we show the results for simulations with $f_\mathrm{PAH}=0.1$ at $\mathrm{3\,Myr}$ along side with lines of constant $M_\mathrm{disc}/\dot{M}_\mathrm{acc,\star}$ (i.e. $\tau_\mathrm{acc}$). The panels in Figure~\ref{fig:mdisc_vs_macc} can be compared directly to Fig.~7 in \cite{Manara2022}. While observations typically lie between $\tau_\mathrm{acc}=10\,\mathrm{Myr}$ and $\tau_\mathrm{acc}=0.1\,\mathrm{Myr}$, our simulations mostly fall in between $\tau_\mathrm{acc}=10\,\mathrm{Myr}$ and $\tau_\mathrm{acc}=1\,\mathrm{Myr}$, with some cases, particularly among the MHD wind-driven models, having $\tau_\mathrm{acc}<1\,\mathrm{Myr}$ towards low disc masses $M_\mathrm{disc}<10^{-3}\ \mathrm{M_\odot}$.

The lack of high accretors $\gtrsim 10^{-8}\mathrm{M_\odot/yr}$ (i.e. short $\tau_\mathrm{acc}$) is a known issue in global models \cite[e.g.][]{Emsenhuber2023,tabone_alma_2025}. This tension can be relaxed by more compact discs (see discussion in \cite{Weder2023}) or evolving torques in the MHD wind-driven scenario \cite[e.g.][]{Tabone2022b,tabone_alma_2025}. Further, our simulations show less spread accretion rates, which could be due to accretion variability, which is found to produce variations around $\mathrm{0.4\,dex}$ \cite[e.g.][]{costigan_temperaments_2014,venuti_mapping_2014}. Disc masses are usually obtained from measuring the sub-millimetre continuum and including the dust evolution (i.e. radial drift) in models leads to more spread and generally lower $\tau_\mathrm{acc}$ \citep{Sellek_MaccMdustCorr_2020}. However, both accretion rate and disc mass measurements are also subject to significant systematic uncertainties. For example, the continuum-inferred dust mass, scaled to ISM dust-to-gas ratios, is usually used as a proxy for the total disc mass. \citet{Longarini2025} recently showed that in massive protoplanetary discs where dynamical mass measurements are possible, the dust-to-gas ratio is considerably lower than the canonical $100$, and exhibits substantial scatter. This is one of many potential sources of variation in the inferred $M_\mathrm{disc}-\dot{M}_\mathrm{acc,\star}$ relation.

Assuming long accretion timescales for weakly irradiated regions $\mathcal{F}_\mathrm{FUV,10G_0}$ (as suggested from the lifetimes of highly irradiated regions $\mathcal{F}_\mathrm{FUV,CygOB2}$, see Section~\ref{subsec:disc_frac_evo}) increases this tension between simulations and observations. Therefore, taking our simulations at face value, they suggest that different sets of initial conditions (i.e. $\tau_\mathrm{acc,0}$ and possibly $f_\mathrm{PAH}$) are needed in order to explain disc lifetimes and $M_\mathrm{disc} - \dot{M}_\mathrm{acc,\star}$ relations in both low and high FUV environments. Our results point towards not only strongly variable and time-dependent but also heterogeneous angular momentum transport between different regions.

This is an apparently surprising result. However, both PAH abundance and accretion timescales may plausibly vary between star forming regions. From an observational perspective PAH abundance for individual irradiated discs has been inferred to be substantially depleted \citep{Vicente2013}. However, it is difficult to compare this to discs in low UV regions where PAH emission is not excited. The accretion timescale is a more easily inferred observationally in such low UV regions, although it relies on both accretion rate and disc mass measurements. This can be challenging in more distant (massive) star forming regions. Thus, observationally we lack the data to anchor a direct comparison of either $f_\mathrm{PAH}$ or $\tau_\mathrm{acc}$, particularly on a population level.

From a theoretical perspective, plausibly $f_\mathrm{PAH}$ may be lower in the outer discs if they are destroyed there by strong UV irradiation \citep{Kamp2011}. Although few if any studies have directly explored how external irradiation may modify angular momentum transport, changes to ionisation and thermal structure in discs may plausibly influence a variety of (magneto-)hydrodynamic processes in discs \citep{Lesur2022}. More speculatively, late stage infall may be a significant driver of stellar accretion \citep{winter_planet_2024,Winter2024c}. In this case, hot, irradiated regions, which would typically have a low ISM gas density, may also host discs with lower accretion rates. These possibilities are speculative, but our results highlight the need for further exploration. 

For the remainder of this work, we consider the best fitting populations with $f_\mathrm{PAH}=0.1$ and $\mathcal{F}_\mathrm{FUV,CygnusOB2}$ with $\tau_\mathrm{acc,0,long}$ and $\mathcal{F}_\mathrm{FUV,10G_0}$ with $\tau_\mathrm{acc,0,short}$ to be our fiducial cases (see Table~\ref{tab:fiducial_cases}).

%TTTTTTTTTTTTTT
\begin{table}[!h]
    \begin{center}
    \caption{Summary of in Section~\ref{subsec:disc_frac_evo} identified fiducial cases that are best able to reproduce disc fraction evolution and $M_\mathrm{disc} - \dot{M}_\mathrm{acc,\star}$ relation in low and highly irradiated regions.}
    \label{tab:fiducial_cases}
    \begin{tabular}{llcl}
    \hline \hline
    Scenario & $\mathcal{F}_\mathrm{FUV}$ & $f_\mathrm{PAH}$ & $\tau_\mathrm{acc,0}$ \\ \hline
    MHD wind-driven & $\mathcal{F}_\mathrm{FUV,10G_0}$ & $0.1$ & $\tau_\mathrm{acc,0,short}$ \\
    & $\mathcal{F}_\mathrm{FUV,CygnusOB2}$ & $0.1$ & $\tau_\mathrm{acc,0,long}$ \\
    Viscous & $\mathcal{F}_\mathrm{FUV,10G_0}$ & $0.1$ & $\tau_\mathrm{acc,0,short}$ \\
    & $\mathcal{F}_\mathrm{FUV,CygnusOB2}$ & $0.1$ & $\tau_\mathrm{acc,0,long}$ \\ \hline
    \end{tabular}
    \end{center}
\end{table}
%TTTTTTTTTTTTTT

%--------------
\subsection{Disc fraction gradients with FUV flux} \label{subsec:cygnusOB2_compar}
%--------------

%FFFFFFFFFFFFFF
\begin{figure*}
    \centering
    \includegraphics[width=\linewidth]{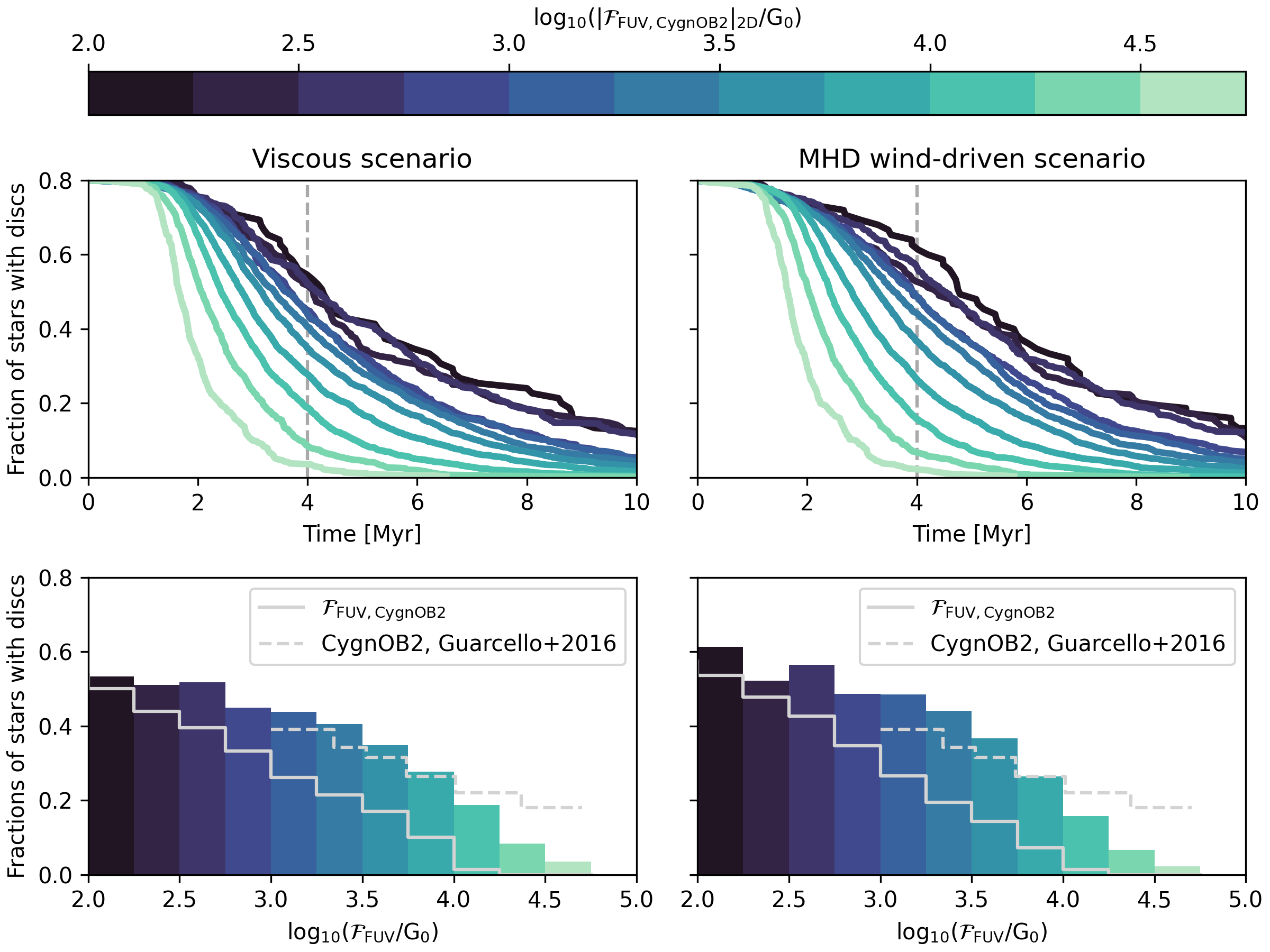}
    \caption{Spatial variation of the disc fraction in a Cygnus OB2-like FUV field environment. The top panel shows the temporal evolution of the disc fraction in the different regions of the 2D projected FUV field for both viscous and MHD wind-driven scenario. The coloured histograms in the bottom panels show as a function of $|\mathcal{F}_{\mathrm{FUV,CygnOB2}}|_\mathrm{2D}$ the fraction of stars with discs at a fixed moment in time of $\mathrm{4\,Myrs}$, indicated in the top panels by the vertical dashed line. The dashed histogram line shows the disc fractions of Cygnus OB2 obtained observationally by \cite{guarcello_photoevaporation_2016}. The solid white line shows the disc fraction in the actual (3D) FUV field in the simulation.}
    \label{fig:disc_frac_cob2_comppar}
\end{figure*}
%FFFFFFFFFFFFFF

\cite{guarcello_photoevaporation_2016} studied the fraction of surviving discs in different parts of the Cygnus OB2 cluster. They used 2D projected distances and calculated the resulting FUV field for the stars in the cluster. They found a smooth decline from 40\% to 18\% of the disc fraction between $\mathrm{10^{3}\,G_0}$ and $\mathrm{10^{4.7}\,G_0}$ - strong evidence of more rapid disc dissipation in regions with intense local UV fields.

Figure \ref{fig:cluster} shows the synthetic cluster with stars coloured by the 2D projected FUV field after star formation has ended. This is a good approximation of Fig.~3 in \cite{guarcello_photoevaporation_2016}. Note that the region where the FUV field exceeds $\mathrm{10^3\,G_0}$ extends for $\mathrm{\sim10\,pc}$. At a distance of $\mathrm{1.5\,kpc}$ this correspond to an angle of $\mathrm{\sim0.7\,deg}$ which is similar to the observed Cygnus OB2 cluster. Figure \ref{fig:disc_frac_cob2_comppar} shows the evolution of the disc fraction inside the different regions of our synthetic cluster for the two evolution scenarios (viscous and MHD wind-driven). The disc fraction is found to be highly dependent on the ambient FUV field strength. While discs in regions with $\gtrsim 10^{4} \mathrm{G_0}$ have $\lesssim 20\%$ discs surviving for more than $\mathrm{4\,Myr}$, regions with $\lesssim 10^{3}\mathrm{G_0}$ retain $\gtrsim 50\%$ of their discs, which is in line with observed disc fractions in low mass clusters (see Fig.~\ref{fig:disc_fraction_evo}). It should be noted that both evolution scenarios show almost identical behaviour.

We find a steady decline of the fraction of stars with discs with increasing FUV field strength, similar to what is seen in Cygnus OB2. The disc fractions are in good agreement for $\mathcal{F}_\mathrm{FUV} \lesssim 10^4\,\mathrm{G_0}$ (down to FUV fluxes contained within the observational data). However, the decline in disc fraction appears to be somewhat steeper with increasing FUV than what is observed. This tension is not as severe since it affects a minority of the sample with high projected fluxes $\mathcal{F}_\mathrm{FUV}>10^{4}\, G_0$. Only $\sim 20$~percent of stars in our distribution have  $\mathcal{F}_\mathrm{FUV}$ greater than this value. These stars also have the smallest separations $x_\mathrm{OB}$ from their dominant UV contributor. Since $\dot{\mathcal{F}}_\mathrm{FUV} \propto \dot{x}_\mathrm{OB}/x_\mathrm{OB}$, these stars also undergo rapid changes in UV flux. If we had included dynamical evolution in our model, we would expect this to flatten the dependence of disc lifetime on FUV flux, as found by \citet{winter_external_2019}. 

We do not expect this overdepletion of highly irradiated discs to substantially alter the disc lifetimes for the population as a whole. If $30$~percent (the fraction at the highest FUV bin at which the model and observations agree) of discs in the highest $20$~percent survive, this would result in a $\sim 6$~percent increase. Even this maximal difference would not substantially change the conclusions inferred from Fig.~\ref{fig:disc_fraction_evo}. In practice the overall difference in disc fraction is probably smaller than this, because dynamical evolution would mean an initially denser (higher FUV) environment, and that some subset of currently lower UV YSOs experienced historically higher ambient UV. Ignoring dynamics is therefore a reasonable simplification in this context.

To illustrate the role of projection effects, the disc fraction as a function of the true 3D FUV field strength is shown as a solid grey line. This illustrates that the number of stars retaining their disc within strong FUV fields is significantly lower and many of the disc residing seemingly in highly irradiated regions are a result of the 2D projection. This is an important consideration when selecting samples of apparently highly irradiated discs that exhibit a NIR excess; these stars probably experience in reality much lower UV flux than apparent from their projected separation from OB stars. This has potential consequences for ALMA and JWST studies \citep[e.g.][]{RamirezTannus2023}, where NIR excess is usually a prerequisite for selection.

%--------------
\subsection{The evolution of disc radii} \label{subsec:disc_radii}
%--------------

%FFFFFFFFFFFFFF
\begin{figure*}
    \centering
    \includegraphics[width=\linewidth]{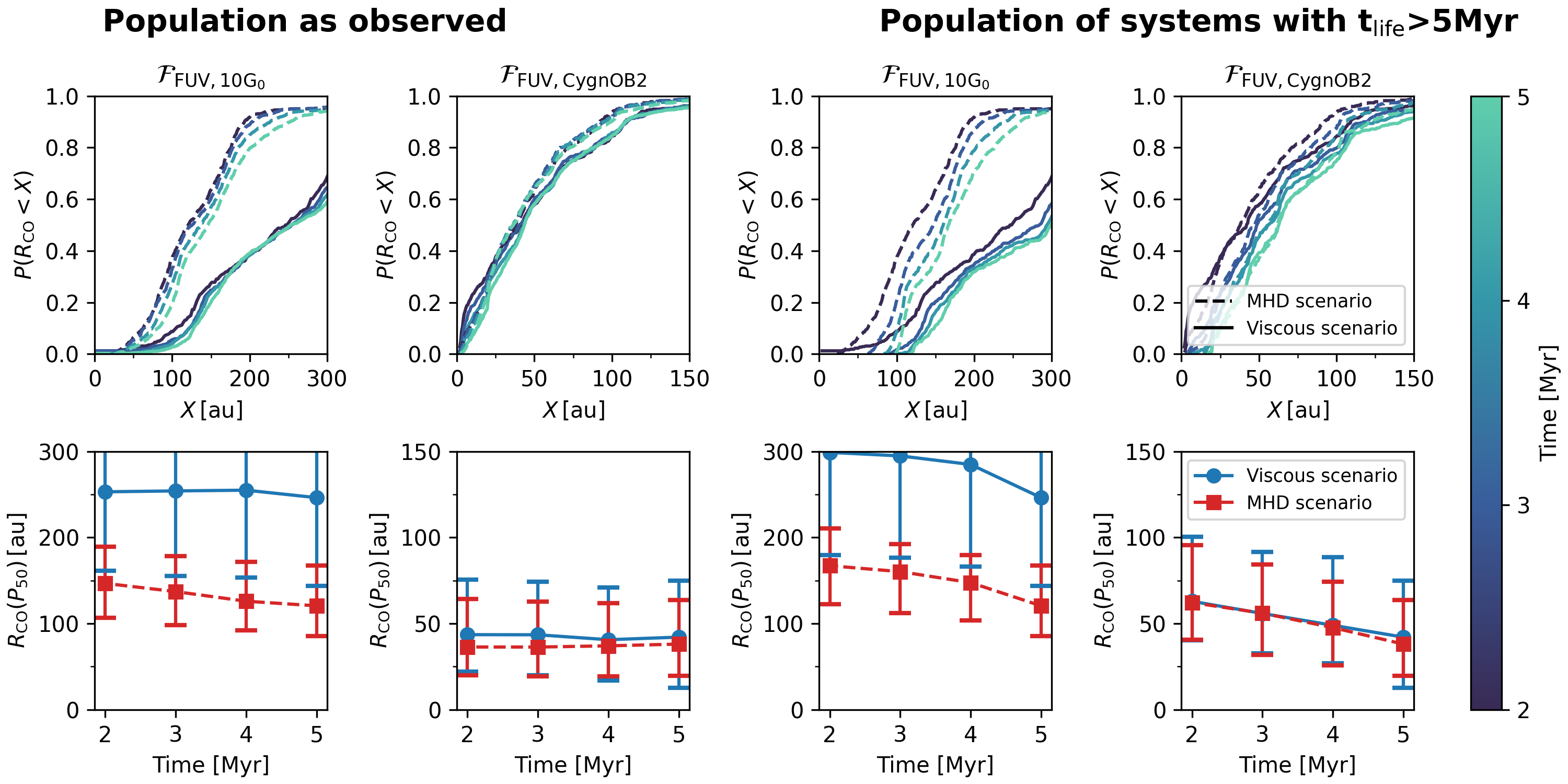}
    \caption{Time evolution of disc radii from fiducial populations (see Table~\ref{tab:fiducial_cases}). The top panels show the time evolution of the cumulative fractions of the disc radii and corresponding bottom panels show the evolution of the 50th percentile of the respective simulations in the top panel with markers indicating the 25th and 75th percentile. Note that the axis limits for the radii differ between low- and high FUV field cases. The left panels show the population as it would be observed (i.e. distribution includes all discs present at the time). The right panels show the same simulations, however only including systems that have a lifetime $t_\mathrm{life}>\mathrm{5\,Myr}$.}
    \label{fig:disc_radii}
\end{figure*}
%FFFFFFFFFFFFFF

Disc radii have been extensively explored as a potential way of distinguishing the MHD versus viscous evolution scenarios. Theory predicts that in the absence of external photoevaporation, discs would expand due to viscosity as opposed to contracting when subject to MHD winds \citep[e.g.][]{Manara2022}. However, \citet{coleman_photoevaporation_2023} and \citet{anania_alma_2025} showed that even low levels of FUV irradiation can obfuscate this effect.

Following \cite{toci_analytical_2023} we define the disc radius using the critical surface density where self-shielding through CO breaks down. The critical $\mathrm{^{12}CO}$ column density is of order $\hat{N}_\mathrm{CO} \simeq 10^{16}\mathrm{cm^{-2}}$. The critical surface density then being given by
\begin{equation} \label{eq:crit_column_density_12CO}
    \Sigma_\mathrm{crit} = \frac{1}{\xi_\mathrm{CO}}\mu_\mathrm{H_2}m_\mathrm{H}\hat{N}_\mathrm{CO},
\end{equation}
where we adopt $\xi_\mathrm{CO}=10^{-5}$ for the relative abundance of $\mathrm{^{12}CO}$ to $\mathrm{H_2}$ with the mean molecular weight of the gas set to $\mu_\mathrm{H_2}=2$ for simplicity. This is in line with observed CO depletion factors \citep[see Fig.~9 in][]{Rosotti2025}. Increasing or decreasing $\xi_\mathrm{CO}$ by one order of magnitude leads to decreased or increased radii (mostly for viscous scenarios with $\mathcal{F}_\mathrm{FUV,10\,G_0}$ due to the less sudden truncation of the outer disc) however has no influence on the findings presented here. For the MHD wind case, and for the high FUV environment of Cygnus OB2, a two order magnitude variation of $\xi_\mathrm{CO}$ nearly has no effect on the inferred radii because of the very steep surface density gradient at the outer edges of the discs.

In Figure~\ref{fig:disc_radii} we show the time evolution of the cumulative fraction for our four fiducial populations (see Table~\ref{tab:fiducial_cases}) and the time evolution of the median, the 25th and the 75th percentile in the corresponding bottom figure. On the left, we show the populations as they would be observed - meaning that at each time, the radius distribution of all discs present at that time are shown. This leaves to an evolving sample of discs, since short lived discs drop out of the sample. Such an effect was also found and discussed by \cite{tabone_alma_2025}, referred to as survivorship bias. On the right, the sample of the population has been selected such that all discs have lifetimes of $t_\mathrm{life}>\mathrm{5\,Myr}$, leading to an initially smaller but steady sample of the population. This selection makes it possible to better understand how individual discs evolve under the impact of the different physical processes, but is not accessible to observations. At 5 Myr, both selections are identical. 

Based on Figure~\ref{fig:disc_radii} we make the following findings (where it is important to distinguish between the evolution of the radii of individual discs and the evolution of the distribution):
\begin{enumerate}[a)]

    \item If the population is taken as it can be observed (i.e. size distribution of all discs present at a given time; left two panels), the evolution of the disc size distribution is limited between 2 and 5 Myr. In the low FUV environment, there is a decrease of the median radius (and the percentiles) of about 25\% for the MHD scenario and no evolution for the viscous scenario. For the high FUV field of Cygnus OB2, the median and percentiles are nearly constant in time for both evolution scenarios. 
    
     \item In contrast, in the two right panels, i.e., when excluding systems with a lifetime $t_\mathrm{life} \lesssim \mathrm{5\,Myr}$ a steady decline of disc sizes is observed for all scenarios. This reflects the evolution of individual discs. In the population as observed, however, this decrease is partially masked, as the initially numerous small discs with low lifetimes decrease the median before 5~Myr. As they disappear, larger discs are left behind. For the FUV field of Cygnus OB2, the size decrease of individual discs is on the population level even stopped and the median remains constant.

    \item Although starting from the same initial conditions (in particular radii), the two disc evolution scenarios produce very distinct size distributions at a given moment in time. In the low FUV case, there is a temporally nearly constant offset of approximately 130~au by which viscous discs are larger. In the Cygnus OB2 high FUV case, discs show almost identical sizes. This can be understood from our discussion in Section~\ref{sec:theoretical_understanding} where we showed that in the viscous scenario discs are decreting at the same rate as they are accreting thus increasing size until external photoevaporation rates balance the decretion rate. However, in our simulations this initial growth phase is very short $\sim10^{5}~\mathrm{yrs}$. This effect is only observed in weakly irradiated regions as for in a strong FUV field, discs cannot spread due to the high external photoevaporation rates.
    
    \item As expected, higher FUV environments lead to clearly more compact discs: for MHD wind-driven evolution, characteristic median sizes are 100-150~au in the $\mathrm{10\,G_0}$ environment versus $\sim$50~au in Cygnus OB2. For viscous evolution, the values are 250~au versus $\sim$50~au.
     
\end{enumerate}

We stress that both scenarios show decreasing or stalling median disc radii even for low FUV field strength of $\mathrm{10\,G_0}$ as opposed to models that do not consider external photoevaporation \cite[see review by][]{Manara2022}. This is in line with previous findings by \cite{coleman_photoevaporation_2023,anania_alma_2025}. 

%FFFFFFFFFFFFFF
\begin{figure}
    \centering
    \includegraphics[width=\linewidth]{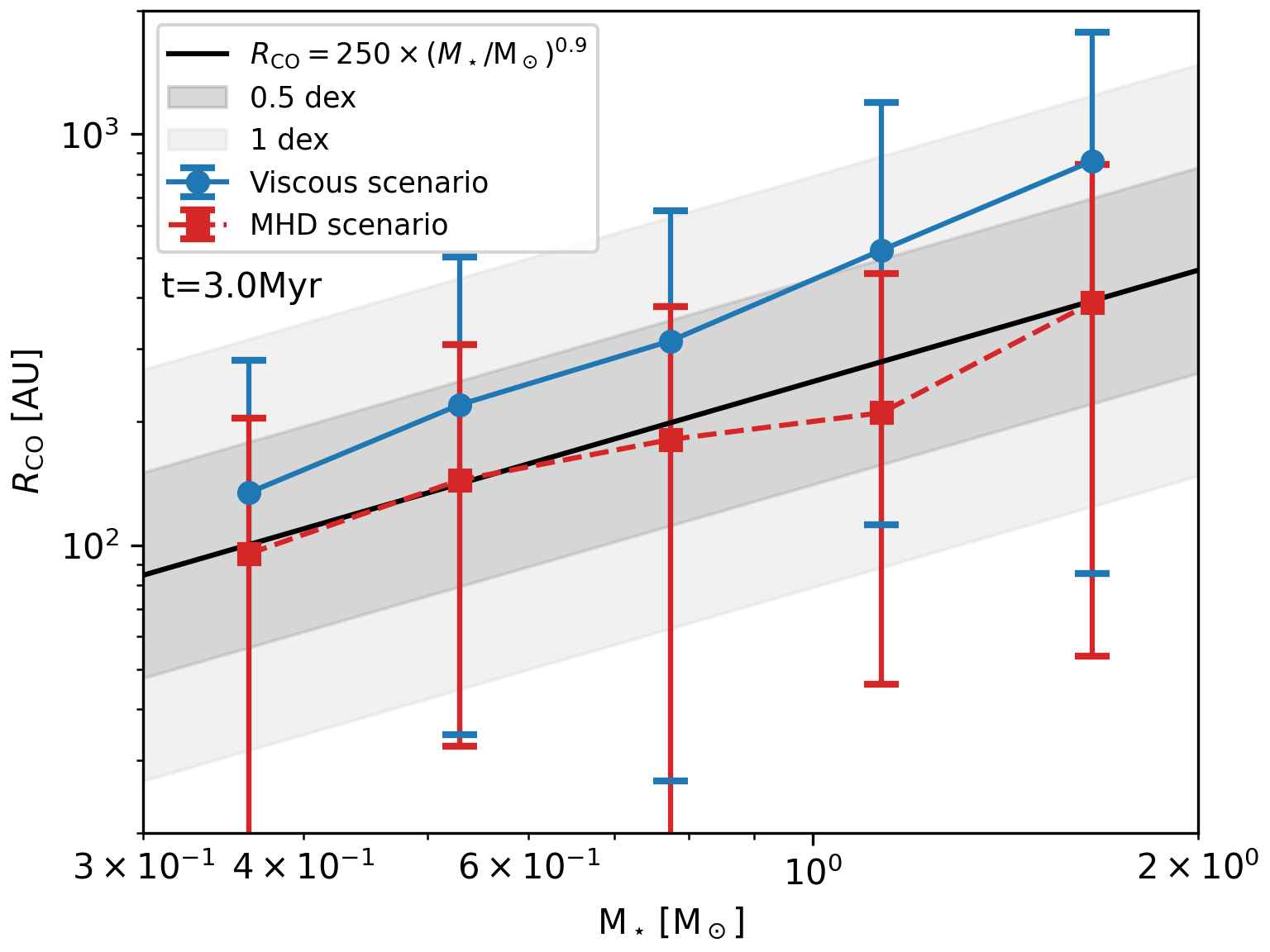}
    \caption{Stellar mass dependence of the disc radii from our fiducial populations with $\mathcal{F}_\mathrm{10\,G_0}$ (see Table~\ref{tab:fiducial_cases}), shown at 3~Myr. The markers indicate the 25th and 75th percentiles. The observational trend (black line) is shown along a 0.5 and 1~dex spread around the mean \citep{Andrews2020}.}
    \label{fig:mstar_rdisc_compar}
\end{figure}
%FFFFFFFFFFFFFF

Radii of many sources have been measured using resolved millimetre continuum measurements resulting in a disc size $R_\mathrm{mm}$. The mm emission however resolves only dust and not the size of the gas disc. Dust sizes can be converted into gas disc sizes using the empirical relation $R_\mathrm{mm}/R_\mathrm{CO} \approx 2.5$ (see Fig.~4 b) in \cite{Andrews2020} and \cite{Sanchis_MeasuringRadiiLupus_2021}. \cite{Andrews2020} compiled a large sample of discs from various nearby star forming regions (see their Fig.~5) and found the scaling between stellar mass and $R_\mathrm{mm}$ to be $R_\mathrm{mm} \propto M_\star^{0.9}$. More recent results from the AGE-PRO survey \citep{zhang_alma_2025} show median disc radii for Lupus and Upper Sco to be 74~au and 110~au, which given the median stellar mass of their sample being $\sim0.4~\mathrm{M}_\odot$ is consistent with \cite{Andrews2020}. The observed disc radii show large spread ($\sim0.5~\mathrm{dex}$ beyond measurement uncertainties). In Fig.~\ref{fig:mstar_rdisc_compar} we show the disc radii dependence on stellar mass for the fiducial simulations with $\mathcal{F}_\mathrm{10\,G_0}$ (see Table~\ref{tab:fiducial_cases}) at 3~Myr. The black line corresponds to the observational trend \citep[Fig.~5 in][]{Andrews2020} with 0.5 and 1~dex around the inferred relationship. Our simulations show good agreement with the observational trend. Disc radii obtained from the MHD evolution scenario are slightly favoured over the viscous evolution scenario.

Although (CO) median outer disc radii are substantially different between the viscous and MHD models (i.e. MHD models predict more compact discs), the large spread makes it challenging to make a reliable direct comparison between models and observations. Continuum radii are difficult to interpret because they are much more compact than the gas \citep[e.g.][]{Andrews2020}, and depend sensitively on opacity variations \citep{Rosotti2019}. CO outer disc radii may depend on freeze-out (and growth/drift of dust) and other chemical effects. Therefore we further consider in the following section how the wind-driven mass-loss rate may offer a novel avenue to distinguish between MHD winds and turbulence.

%--------------
\subsection{Distinguishing MHD winds from viscous evolution: a novel approach via external mass-loss rate versus stellar accretion rate} \label{subsec:distinguish_mhd_and_viscous}
%--------------

%FFFFFFFFFFFFFF
\begin{figure*}
    \centering
    \includegraphics[width=\linewidth]{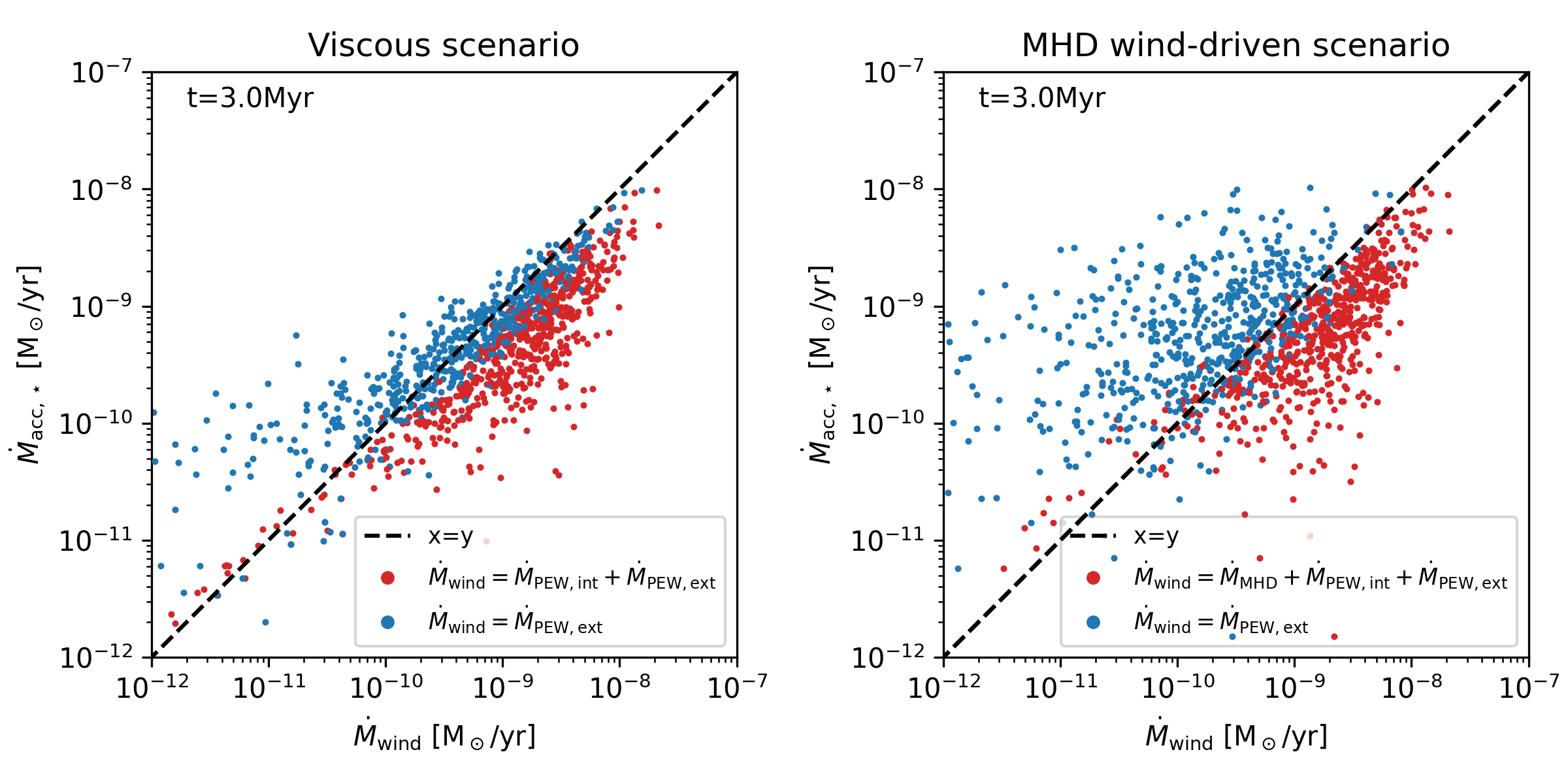}
    \caption{Stellar accretion rate $\dot{M}_\mathrm{acc,\star}$ versus outflow rate $\dot{M}_\mathrm{wind}$ for our fiducial simulations with $\mathcal{F}_\mathrm{FUV,CygnOB2}$ (see Table~\ref{tab:fiducial_cases}), shown at $\mathrm{3\,Myr}$. Red dots take the full mass outflow due to internal and external photoevaporation and an MHD wind into account ($\dot{M}_\mathrm{wind}=\dot{M}_\mathrm{MHD}+\dot{M}_\mathrm{PEW,int}+\dot{M}_\mathrm{PEW,ext}$). Blue dots correspond to taking only the outflow due to external photoevaporation into account ($\dot{M}_\mathrm{wind}=\dot{M}_\mathrm{PEW,ext}$). The left panel shows the results obtained for the viscous scenario and the right panel shows the results obtained for the MHD wind-driven scenario.}
    \label{fig:distinguishing_scenario}
\end{figure*}
%FFFFFFFFFFFFFF

As noted by several authors, for a viscously expanding disc, externally driven wind mass-loss should eventually balance stellar accretion rate \citep{Clarke2007, Winter2020a, Hasegawa2022}. This needs not be the case for MHD wind driven accretion, and therefore comparing accretion rate to wind mass-loss offers a potential way to distinguish between angular momentum transport mechanisms. However, as we will show below, this critically depends on whether the externally driven wind can be distinguished from the internally driven winds.

In Figure \ref{fig:distinguishing_scenario} we show accretion rates and wind mass-loss from our fiducial simulations with a strong FUV field $\mathcal{F}_\mathrm{FUV,CygnOB2}$ (Table~\ref{tab:fiducial_cases}) at $\mathrm{3\,Myr}$, analogous to the theoretical expectations sketched in Figure \ref{fig:scenario_overview}. These figures show that after 3~Myr, when all contributions (internal and external) to the wind are considered (red dots), both MHD and viscous disc populations have $\dot{M}_\mathrm{wind} \gtrsim \dot{M}_\mathrm{acc,\star}$. Note that many cases show lower accretion rates compared to the wind mass-loss rate which is due high internal photoevaporation rates of $\sim 3.93 \cdot 10^{-9}(M_\star/\mathrm{M_\odot})\,\mathrm{M_\odot/yr}$ adding to the total wind mass-loss rate. However, when considering the mass-loss rate due to external photoevaporation only, viscous and MHD wind-driven scenarios show substantial differences (blue dots). While $\dot{M}_\mathrm{acc,\star}$ and $\dot{M}_\mathrm{PEW,ext}$ remain correlated down to a few $10^{-10}\mathrm{M_\odot/yr}$ for the viscous case, in the MHD wind-driven scenario the externally driven wind mass-loss rate drops well below the accretion rate $\dot{M}_\mathrm{wind,ext} \ll \dot{M}_\mathrm{acc,\star}$ throughout the majority of the sample.
In other words: if turbulent viscosity is the main driver of disk evolution, stellar accretion rate and external photoevaporation rate should be comparable for all discs. In the MHD-wind picture, external photoevaporation rates are in contrast typically smaller (up to 3 orders of magnitude) than the stellar accretion rate. This holds for stellar accretion rates larger than a few $10^{-10}\mathrm{M_\odot/yr}$, below which internal photoevaporation starts to play a role as well. Since mass-loss profiles for internal X-ray photoevaporation extend to large radii, they eventually overlap with the region where external photoevaporation takes place in compact discs, loosening the correlation of $\dot{M}_\mathrm{acc,\star}\sim \dot{M}_\mathrm{PEW,ext}$ once the accretion rate (and therefore the decretion rate) significantly drops below the internal photoevaporation rate. However for stars accreting at high rates the difference between the viscous and MHD wind-driven scenario remains clearly visible.
Thus, measuring the externally driven mass-loss rate individually (separate from other winds) would therefore be a critical step and new approach in differentiating angular momentum transport mechanisms. 

This raises the question of feasibility in isolating the externally driven wind observationally. The simplest method for estimating mass-loss in proplyds is by measuring the ionisation front radius \citep[e.g.][]{Henney1998, Vicente2005, Aru2024}. However, this is a measure of recombination balance and hence the total wind-driven mass-loss. It is also only possible for the brightest ionisation fronts. Nonetheless, there are some promising ways of estimating mass-loss rates using alternative tracers \citep{PlanetFormationEnvironments2025}. This may involve utilising emission lines that trace hot gas at the base of the wind, such as [OI] \citep{Ballabio2023} or [CI] \citep{Aru2024}. Alternatively, if $f_\mathrm{PAH}$ can be reliably constrained or controlled for, mass-loss rates based on observed disc properties and the FRIED grid could also be used. This would require concerted efforts to characterise disc properties and accretion rates among a large sample of irradiated discs, but this may be an achievable goal in the near future.

%===============
\section{Summary and conclusions} \label{sec:summary_conclusion}
%===============
Whether turbulent viscosity or MHD winds dominate disc evolution is a fundamental open question. In this work, we explored how these two evolutionary pathways manifest in the observational properties of discs in irradiated star forming regions.

We ran comparative synthetic disc populations for both evolution scenarios, both including internal X-ray and external FUV photoevaporation. We varied the accretion timescales $\tau_\mathrm{acc}$ for both weakly irradiated regions ($\mathcal{F}_\mathrm{FUV,10G_0}$) and strongly irradiated regions ($\mathcal{F}_\mathrm{FUV,CygnOB2}$), where for the latter we used a self-consistent approach to calculate the FUV field in a cluster similar to Cygnus OB2, and varied the polycyclic aromatic hydrocarbon abundance $f_\mathrm{PAH}$ that influences the external photoevaporation rates. We then compared the time evolution of the disc fraction with observed disc fractions from representative clusters and investigated disc mass and stellar accretion rate relations, identifying sets of parameters that are able to reproduce the observed trends. We then further verified our models by comparing the spatial variation of disc fraction in the highly irradiated cases to observations from Cygnus OB2. We further show the evolution of disc radii and put our hypotheses of distinguishing between viscous and MHD wind-driven disc evolution using the outflow rates to the test.

We present the following findings based on our model:
\begin{enumerate}
    \item We find fundamental differences in evolution of low and high mass clusters that are not explained by the FUV field strength only. Our simulations suggest that the initial accretion timescales $\tau_\mathrm{acc,0}$ and thus the efficiency of angular momentum transport / extraction with associated $\overline{\alpha_{r\phi}}$ and $\overline{\alpha_{\phi z}}$ distributions might differ between different star forming regions.

    \item At the (observable) population level, we find limited temporal evolution of the median outer disc radii between 1 and 5~Myr both in the viscous and MHD wind paradigm. For a low FUV environment, median radii decrease by about 25\% for MHD wind-driven discs or stall for viscous discs, while in a high Cygnus OB2-like environment, the radii are approximately constant for both evolution scenarios.
    
    \item At a given moment in time, however, the viscous evolution scenario predicts larger median disc radii than MHD wind-driven evolution in low FUV fields (about 100~au difference).

    \item Outer disc radii are strongly influenced by external photoevaporation. For MHD wind-driven evolution, median radii are $\sim$120~au at $10\,\mathrm{G_0}$ versus $\sim$50~au in Cygnus OB2. For viscous evolution, radii are $\sim$250~au versus $\sim$50~au.

    \item At the level of the temporal evolution of individual long-lived discs (which cannot be observed directly) the outer radii contract, sometimes substantially. This is true regardless of whether a disc evolves viscously or under MHD winds, and whether it is in a low $10\,\mathrm{G_0}$ environment or in Cygnus OB2. 
        
    \item Our simulations show that the mass-loss-rates due to external photoevaporation in the MHD wind-driven scenario are usually much lower (up three orders of magnitude) than the accretion rate onto the star, whereas for the viscous scenario they are closely related (differences of less than $\sim$50\%) as expected from angular momentum conservation. If observations can disentangle internally and externally driven winds in irradiated regions, then this offers a novel pathway to establish whether MHD-winds or turbulent viscosity drive disc evolution. 
\end{enumerate}

While current observational constraints can be explained by both evolution scenarios our findings point out a fundamental difference between disc evolution in low and highly irradiated regions that is independent of the disc evolution scenario that has still to be understood.

This work paves the way towards modelling planet formation in specific environments for which fundamental open questions exist \citep{mordasiniburn2024} and thus assess the impact of the cluster on the emerging planet population, which will be investigated in future works.

% ----------------
% acknowledgements
% ----------------
\begin{acknowledgements}
First, we would like to thank Mario Guarcello for providing additional insights on the Cygnus OB2 sample. We thank also Thomas Haworth, Richard Nelson, Gavin Coleman and Alexandre Emsenhuber for stimulating discussions on this work. We further thank the anonymous referee for a thorough read and helpful comments that improved the manuscript. J.W. and C.M. acknowledge the support from the Swiss National Science Foundation under grant 200021\_204847 “PlanetsInTime”. Part of this work has been carried out within the framework of the NCCR PlanetS supported by the Swiss National Science Foundation under grants 51NF40\_182901 and 51NF40\_205606. Calculations were performed on the Horus cluster of the Division of Space Research and Planetary Sciences at the  University of Bern.  AJW has been supported by the European Union’s Horizon 2020
research and innovation programme Marie Skłodowska-Curie grant agreement No 101104656 and by the Royal Society through a University Research Fellowship, grant number URF\textbackslash R1\textbackslash 241791.
\end{acknowledgements}

\bibliographystyle{bibtex/aa.bst}
\bibliography{references.bib}

\onecolumn
\begin{appendix}
    %===============
    \section{Collection of representative clusters} \label{app:observed_disc_fractions}
    %===============
    Disc lifetimes are inferred from comparing the fraction of stars having a disc (i.e. showing infra red emission excess) from clusters of different ages dating back to the work of \cite{HaischJr.2001}. Whereas the fraction of discs are usually well constrained (given one can associate the members of a cluster), ages are far less well constrained partly because clusters do not form instantaneously but also because determining ages from stellar evolution tracks remains hard and depends on the models used \cite[e.g.][]{siess_internet_2000,Baraffe2015}.

    Further, the lifetime is linked to the cluster environment (e.g. the presence OB type stars) \cite[i.e.][]{michel_bridging_2021,Pfalzner_discFraction_2022}. We therefore compare our simulations to a selection of clusters that are representative for the corresponding simulations. The selected clusters are listed in Table \ref{tab:disc_fractions}.
    
    %TTTTTTTTTTTTTT
    \begin{table*}[!h]
        \begin{center}
        \caption{Selected sample of clusters used in Fig. \ref{fig:disc_fraction_evo}.}
        \label{tab:disc_fractions}
        \renewcommand{\arraystretch}{1.5}
            \begin{tabular}{lccccc}
                \hline\hline
                & Region & $\mathcal{F}_\mathrm{FUV,median}$\textsuperscript{(1)} & Age\textsuperscript{(2)} & Disc fraction\textsuperscript{(3)} & References \\ 
                & & $\mathrm{[G_0]}$ & [Myr] & [\%]\\
                \hline
                \multicolumn{3}{l}{Low mass regions:} $\mathcal{F}_\mathrm{FUV} \leq 20\,\mathrm{G_0}$ \\
                \hline
                & UppSco        & $20.11_{-10.7}^{+21.03}$      & $8.5_{-3.5}^{+3.5}$   & $19_{-1}^{+1}$        & (a),(f),(f)   \\ 
                & Lupus         & $3.48_{-0.31}^{+21.03}$       & $2.0_{-0.7}^{+1.6}$   & $50_{-5}^{+5}$        & (a),(h),(c)   \\ 
                & Taurus        & $2.92_{-0.14}^{+0.16}$        & $0.9_{-0.4}^{+0.8}$   & $64_{-5}^{+5}$        & (a),(h),(g)   \\ 
                & Cham I        & $4.57_{-1}^{+4.3}$            & $2.8_{-1.4}^{+3.8}$   & $44_{-4}^{+4}$        & (a),(h),(d)   \\ 
                & Cham II       & $2.76_{-0.02}^{+0.03}$        & $1.7_{-1}^{+1}$\textsuperscript{\dag}         & $76_{-14}^{+14}$      & (a),(d),(d)   \\ 
                & CrA           & $9.94_{-5.69}^{+14.83}$       & $5.5_{-1}^{+1}$\textsuperscript{\dag}         & $18_{-1}^{+1}$        & (a),(b),(b)   \\ 
                \hline
                \multicolumn{3}{l}{High mass regions:} $\mathcal{F}_\mathrm{FUV}>100\,\mathrm{G_0}$ \\
                \hline
                & $\sigma$ Ori  & $232.54_{-103.7}^{+946.43}$   & $4_{-1}^{+1}$         & $36_{-4}^{+4}$        & (a),(l),(k)   \\ 
                & $\lambda$ Ori & $536.86_{-436.5}^{+2602.54}$  & $5_{-1}^{+1}$         & $18.5_{-4}^{+4}$      & (a),(j),(j)   \\ 
                & Cygnus OB2    & $1806_{-1166}^{+3046}$         & $4_{-1}^{+1}$         & $33_{-1}^{+1}$          & (m),(i),(e)   \\ 
                \hline
            \end{tabular}
        \end{center}
        Notes: (1) median FUV fluxes in the cluster with uncertainties (16\textsuperscript{th} and 84\textsuperscript{th} percentiles of the distribution), (2) ages compiled from various references with uncertainties reflecting the age spread in the cluster, (3) disc fractions with standard errors compiled from various sources. \\ \dag No information on the age spread is provided and a spread of 2\,Myr is assumed. \\
        References: (a) \cite{anania_novel_2025}, (b) \cite{galli_corona-australis_2020}, (c) \cite{galli_lupus_2020}, (d) \cite{galli_chamaeleon_2021}, (e) \cite{guarcello_photoevaporation_2016}, (f) \cite{luhman_census_2022}, (g) \cite{manzo-martinez_evolution_2020}, (h) \cite{Testi2022a}, (i) \cite{wright_massive_2015}, (j) \cite{hernandez_spitzer_2010}, (k) \cite{hernandez_spitzer_2007}, (l) \cite{oliveira_l-band_2004}, (m) this work; Fig. \ref{fig:init_cond}.
    \end{table*}
    %TTTTTTTTTTTTTT
    \newpage
    %===============
    \section{Initial conditions for the populations} \label{app:initial_conditions}
    %===============
    %FFFFFFFFFFFFFF
    \begin{figure*}[ht!]
        \includegraphics[width=\linewidth]{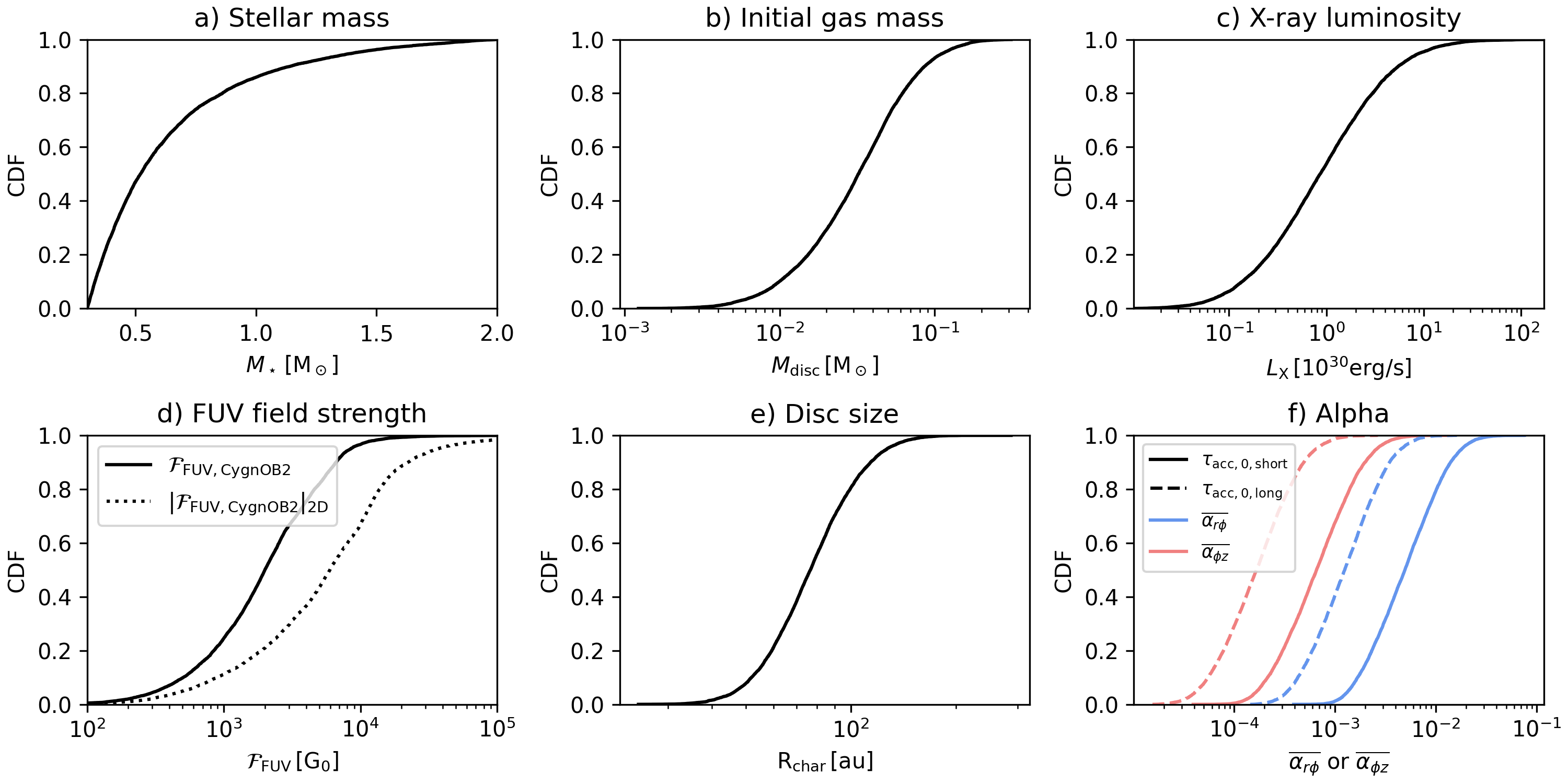}
        \centering
        \caption{Distributions of the initial conditions to set up our synthetic population, described in Section~\ref{subsec:init_cond}. Note that panel d) shows both the distribution of the actual 3D FUV field experienced in the cluster as well as the 2D projection and the X-ray luminosity distribution is shown in panel c). In panel f) we show the two sets of $\overline{\alpha_{r\phi}}$ and $\overline{\alpha_{\phi z}}$ that result from the two accretion timescales adopted (see Figure~\ref{fig:acc_timescale}).}
        \label{fig:init_cond}
    \end{figure*}
    %FFFFFFFFFFFFFF

\end{appendix}

\end{document}